\documentclass[sigconf]{acmart} %
\AtBeginDocument{%
  }

\setcopyright{acmlicensed} %
\copyrightyear{2026}
\acmYear{2026}
\setcopyright{cc}
\setcctype{by}
\acmConference[CCS '26]{Proceedings of the 2026 ACM SIGSAC Conference on Computer and Communications Security}{November 15--19, 2026}{The Hague, Netherlands}
\acmBooktitle{Proceedings of the 2026 ACM SIGSAC Conference on Computer and Communications Security (CCS '26), November 15--19, 2026, The Hague, Netherlands}
\acmDOI{10.1145/3830454.3846627}
\acmISBN{979-8-4007-2871-6/2026/11}  %

\usepackage{graphicx}
\usepackage{subcaption}
\usepackage{array}
\usepackage{xspace,soul}
\usepackage{multirow}
\usepackage[
  separate-uncertainty = true,
  multi-part-units = repeat
]{siunitx}
\usepackage{algorithm}
\usepackage{anyfontsize}
\usepackage[noend]{algpseudocode}
\usepackage{xurl}

\newcommand{\eg}{e.g.\@\xspace}
\newcommand{\ie}{i.e.\@\xspace}
\newcommand{\etal}{et~al.\@\xspace}
\newcommand{\US}{U.\kern0.5ptS.\@\xspace}

\usepackage[most]{tcolorbox}
\usepackage{xcolor}

\newcommand{\myparagraph}[1]{\smallskip\noindent\textbf{\textit{#1}}\hspace{3pt}}

\newcounter{takeaway}

\newcommand{\takeaway}[1]{%
  \refstepcounter{takeaway}%
  \begin{tcolorbox}[
    breakable,
    colback=gray!12,
    colframe=gray!45,
    boxrule=0.4pt,
    arc=2pt,
    left=6pt,right=6pt,top=5pt,bottom=5pt,
    width=\columnwidth,
    before skip=6pt,
    after skip=6pt
  ]
  \textbf{Takeaway \thetakeaway.} #1
  \end{tcolorbox}
}

\usepackage{enumitem}
\newenvironment{wideitemize}
  {\begin{itemize}[leftmargin=*, itemsep=0.4em, topsep=0.4em]}
  {\end{itemize}}

\usepackage{nicematrix}
\usepackage{makecell} %
\usepackage[table]{xcolor}

\begin{document}

\title[Evaluating Practical Enumeration and Blocking Attacks on the Snowflake Circumvention System]{Evaluating Practical Enumeration and Blocking Attacks on the Snowflake Circumvention System}

\author{Linden Chen}
\email{linden@ucsc.edu}
\affiliation{%
   \institution{University of California, Santa Cruz}
   \city{Santa Cruz}
   \state{California}
   \country{USA}
 }
\author{Ryan Sangha}
\email{rysangha@ucsc.edu}
\affiliation{%
   \institution{University of California, Santa Cruz}
    \city{Santa Cruz}
   \state{California}
   \country{USA}
 }
\author{Cecylia Bocovich}
\email{cohosh@torproject.org}
\affiliation{%
   \institution{The Tor Project}
   \city{New York}
   \state{New York}
   \country{USA}
 }
\author{Ram Sundara Raman}
\email{rsundar2@ucsc.edu}
\affiliation{%
   \institution{University of California, Santa Cruz}
    \city{Santa Cruz}
   \state{California}
   \country{USA}
 }

\begin{abstract}
Proxy-based Internet censorship circumvention tools like Snowflake rely on large, dynamic pools of third-party proxies to resist IP-based blocking. We focus on two assumptions underpinning the security of Snowflake: that adversaries cannot easily enumerate proxy IPs, and that blocking those proxies would incur unacceptable collateral damage. In this paper, we test these assumptions by studying practical enumeration and blocking attacks against Snowflake conducted by malicious clients. 

We combine bounded, ethical real-world measurements with large-scale simulation to evaluate both present-day enumeration and blocking risk and broader attacker capabilities. Over 48 days of real-world measurements from May--June 2025, our attack enumerated over 21,000 unique proxy IP addresses belonging to almost 1,000 autonomous systems. Despite this high number, we find that proxy churn limits the overall effectiveness of enumeration over time, and reduces the impact on clients of individual proxy addresses being blocked. However, at the network level, blocking the top 1\% of observed autonomous systems blocks more than 30\% of observed Snowflakes while affecting 0\% of Tranco Top 100 domains and $\sim$2.5\% of Top 1M domains. We discover that the broker's load-aware matching reveals stable, high-capacity proxies to attackers early, especially during periods of elevated demand such as the censorship even in Iran of June 2025, subsequently exposing the networks that contribute disproportionately to system connectivity. In simulation, increasing attacker scale sharply improves both enumeration and blocking success, while higher proxy churn significantly reduces blocking effectiveness. We conclude by discussing and evaluating practical mitigations, some of which have been integrated into Snowflake.

\end{abstract}



\maketitle

\section{Introduction}
\label{sec:intro}
Nation-state actors are increasingly monitoring and censoring Internet traffic, imposing growing restrictions on online content through network middleboxes that detect and block access to specific websites and IP addresses~\cite{freedom-net-2024,aryan2013internet,how-great-gfw,sundararaman2020censoredplanet,ooni}. In response, the Internet freedom community has developed censorship circumvention tools that obscure the true destination of user traffic. Among these, proxy-based solutions such as Snowflake~\cite{bocovich2024snowflake} and VPNs~\cite{ramesh2023all} route traffic through third-party hosts beyond the censor's control, and serve tens of thousands of users daily~\cite{snowflake-broker-metrics,maghsoudlou2023characterizing}. The censorship resistance of these proxy-based circumvention tools depends in part on two principles: (i) \textit{enumeration resistance}: an adversary cannot comprehensively list (and block) all IP addresses associated with the service, and (ii) \textit{collateral blocking}: attempts to block the proxy risk affecting widely used or benign services hosted on the same IP addresses~\cite{bocovich2024snowflake,fifield2015blocking,tschantz2016sok}. 

Recent events suggest that these blocking-resistant properties deserve renewed empirical scrutiny. Iran’s efforts to build a national Intranet~\cite{iran-intranet} and Russia’s willingness to block large portions of the network infrastructure~\cite{ramesh2019decentralized,russia-throttling-cloudflare} show that nation-state censors may choose to perform aggressive blocking, even at substantial collateral cost, during periods of political unrest~\cite{shutdowns}.  
These capabilities raise an important practical question: \textit{how resistant are proxy-based circumvention tools like Snowflake to IP enumeration and blocking attacks in practice?} While many analyses of circumvention tool resilience focus on identifiable traffic characteristics~\cite{xue2022openvpn,gfw-fully-encrypted,wails2024precisely,xue2025discriminative,fifield2016fingerprintability}, they do not empirically test resistance against active enumeration and blocking attacks by censors posing as legitimate clients.

In this paper, we evaluate the practical enumeration and blocking resistance of Snowflake, a prominent circumvention system and widely used pluggable transport for the Tor anonymity network~\cite{bocovich2024snowflake}. Snowflake is designed to resist blocking through scale: it connects censored clients to a large, dynamic pool of lightweight WebRTC proxies (``snowflakes'') through a rendezvous step facilitated by a centralized service called a broker.
We ask whether malicious clients can repeatedly poll the live broker to enumerate proxies and then use that visibility to support IP and network-level blocking. Answering this question empirically is challenging: real-world measurements are necessary to capture broker behavior, proxy churn, and demand-driven effects in deployment, but aggressive probing against a production circumvention system raises clear ethical and operational risks. We therefore combine bounded, ethical real-world measurements with large-scale controlled simulation, which allows us to both characterize a practical attacker's view of a highly dynamic and diverse pool of over 100,000 volunteer circumvention proxies and measure how attack success changes across broader adversarial settings. Our approach differs from prior theoretical studies of optimal proxy distribution by evaluating attacks against a deployed system under real-world constraints~\cite{nasr2019enemy,kon2024spotproxy,fares2026game}. The idea of querying a proxy-distribution service to discover endpoints is not itself new; related attacks have previously targeted Tor bridge distribution~\cite{ling2013tor}. Our key contribution is to determine empirically how this attack operates against the deployed Snowflake ecosystem, where proxies are substantially more numerous, ephemeral, heterogeneous, and selected through a load-aware centralized broker. We further evaluate whether partial enumeration provides sufficient visibility for effective IP- and network-level blocking.

Our real-world longitudinal enumeration experiment involves two low-rate malicious clients polling the live Snowflake broker over 48 days in May--June 2025. These measurements were developed through detailed discussion with the Tor Research Safety Board to ensure they would neither disrupt the live circumvention system nor expose sensitive information. Our probing enumerated more than 21,000 unique proxy IP addresses across almost 1,000 Autonomous Systems (ASes), revealing insightful patterns about how enumeration behaves with proxy churn and changes in demand.
We complement these real-world measurements with a Snowflake-specific simulator that instantiates lightweight Snowflake proxies and clients interacting with a local deployment of the unmodified Snowflake broker, allowing month-long experiments at 100\% system scale. We use the simulator to quantify how attacker scale, Snowflake churn, and connection duration affect enumeration coverage and blocking impact, and to evaluate candidate defenses.  

We make three primary contributions:
\begin{wideitemize}
\item \textbf{Empirical real-world evaluation of a deployed circumvention tool}. We present a systematic, real-world study of practical enumeration and blocking attacks against a live large-scale proxy-based circumvention system. With modest malicious probing, we enumerate over 21{,}000 proxy IPs and show the broker's load-aware matching surfaces a stable, high-capacity subset of proxies with disproportionate blocking value, though enumeration is limited by proxy churn. We find that Snowflake deployments are concentrated in residential access networks, making AS-level blocking highly disruptive with limited web collateral: blocking the top 1\% of observed ASes disrupts over 30\% of observed proxies and affects no Tranco Top~100 domains and only 2.45\% of Tranco Top~1M domains~\cite{tranco}. We also show that elevated demand during the June 2025 Iran shutdown shifted attacker visibility toward a smaller, more stable, and more infrastructure-heavy subset of proxies~\cite{snowflake-iran-june-2025,net4people-iran-june-2025,cui2026characterizing}.

\item \textbf{Full-scale controlled analysis of enumeration and blocking.} We build and open-source a reusable Snowflake simulation framework that runs the Snowflake broker code with lightweight client and proxy abstractions, enabling repeated 30-day experiments at full network scale. We use it to isolate how attacker scale, proxy churn, connection duration, and proxy type affect enumeration coverage and client retry burden. 

\item \textbf{Security implications and actionable mitigations.} We quantify how malicious proxies can attract a disproportionate share of client assignments, and discuss defenses including rate limiting and broker-scheduled proxy polling. We responsibly disclosed these findings and worked with Snowflake maintainers on corresponding mitigations.
\end{wideitemize}

Overall, we show that Snowflake is not trivially enumerable, but a practical attacker can quickly identify the subset of networks that host the majority of proxies, enabling network-level disruption with relatively low collateral cost. More broadly, the paper demonstrates how bounded live measurements and controlled simulation can be combined to evaluate deployed circumvention systems without subjecting their users to aggressive experimentation. These findings are likely not unique to Snowflake and may extend to other volunteer-based proxy systems such as Tor Bridges, Lantern and Psiphon~\cite{lantern, psiphon-webpage}. We propose and evaluate practical mitigations, including client-side rate limits and broker-scheduled proxy polling. We have worked with Snowflake maintainers to implement patches for the latter, which Snowflake maintainers have progressively integrated into the codebase [\textit{Reference Blinded}]. We hope our study provides a pathway for future work aiming to conduct ethical studies of circumvention systems, and leads to more resilient proxy-based circumvention tools. 

\section{Background}
\label{sec:background}
We summarize Internet censorship and circumvention mechanisms in \S\ref{sec:censorship}, before describing Snowflake's design in detail in \S\ref{sec:snowflake-background}. We summarize related work later in \S\ref{sec:related-work}.  

\subsection{Internet Censorship \& Circumvention}
\label{sec:censorship}
Internet censorship, typically performed by nation-state intermediaries, can be implemented through various methods such as DNS manipulation~\cite{pearce2017global,tsai2023detecting}, IP blocking~\cite{ramesh2019decentralized,pearceaugur}, and HTTP(S) or other application-layer blocking~\cite{sundararaman2020measuring,ververis2020cross}. 

Circumvention tools are critical for ensuring open Internet access during these periods of unrest~\cite{ooni-mahsa-amini,tschantz2016sok}. These tools generally employ a combination of techniques, including: (1) \textit{proxy-based routing}, where traffic is routed through an unblocked third party outside the censor’s control~\cite{bocovich2024snowflake,dingledine2004tor,nasr2020massbrowser,kon2024spotproxy}, (2) \textit{steganography}, where censored traffic is disguised within permitted traffic patterns~\cite{mohajeri2012skypemorph,fifield2015blocking}, and (3) \textit{packet manipulation}, where blocked traffic is manipulated at the network or application layer such that deep packet inspection mechanisms fail to correctly classify the traffic~\cite{bock2019geneva,wang2020symtcp}. In this work, we focus on Snowflake, a popular circumvention tool that combines ephemeral, volunteer-operated proxies with traffic obfuscation to evade network-level blocking~\cite{bocovich2024snowflake}. %

\subsection{Snowflake}
\label{sec:snowflake-background}
\begin{figure}
      \includegraphics[width=\linewidth]{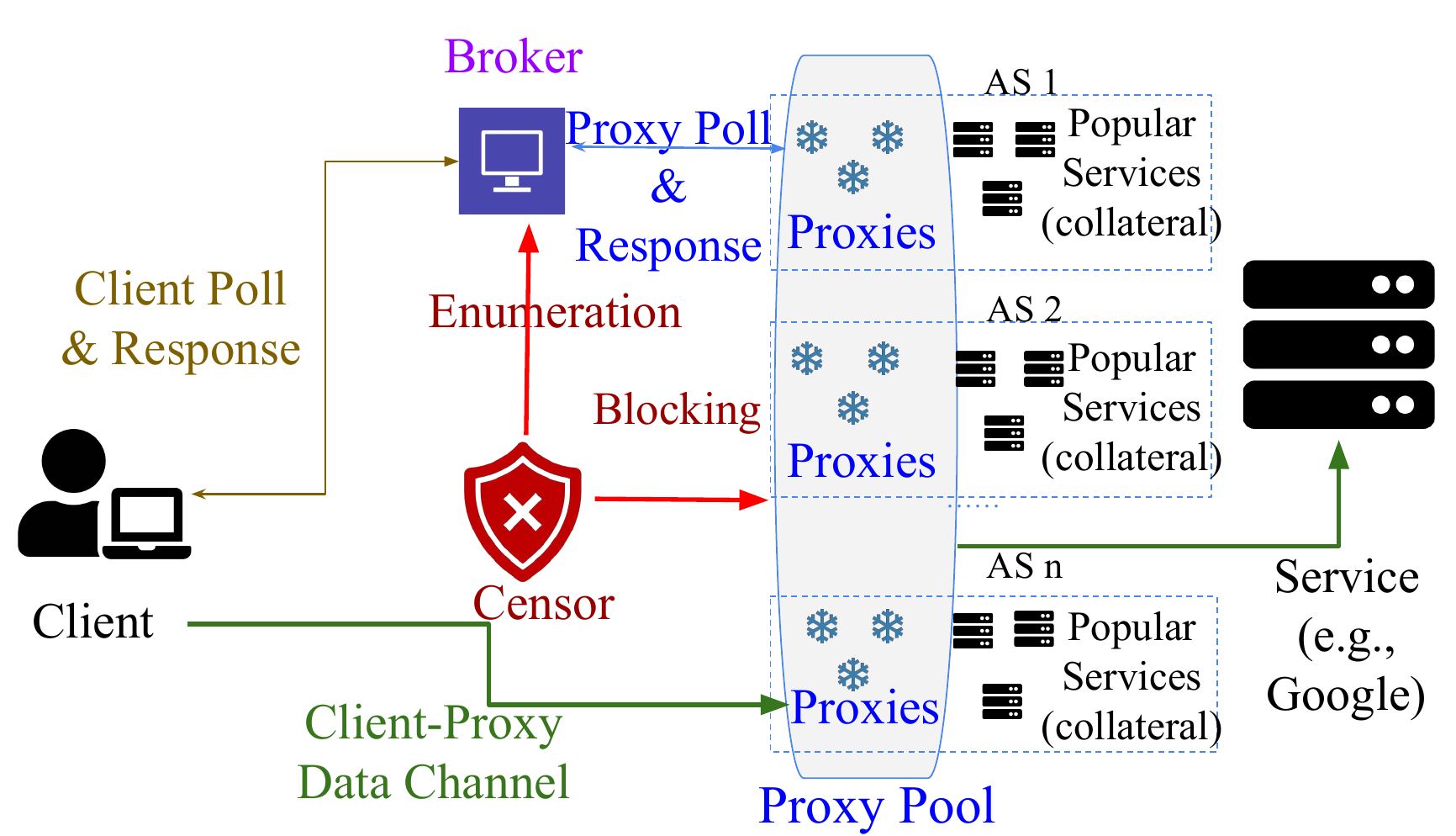}
      \caption{Overview of Enumeration and Blocking Attacks. The figure shows the major participants in the Snowflake circumvention ecosystem, highlighting the targets of the enumeration (broker) and blocking (proxies) attacks.}
      \label{fig:attack-overview}
    \end{figure}

Snowflake is a popular proxy-based censorship circumvention system~\cite{bocovich2024snowflake}. It relies on a large pool of lightweight volunteer-run proxies that relay traffic between clients and the Tor anonymity network from within a censored region. %
Snowflake's circumvention operates in three main stages as illustrated in Figure~\ref{fig:attack-overview} (and Figure 1 in~\cite{bocovich2024snowflake}): \textit{rendezvous}, where clients are matched with available proxies and connection establishment information is exchanged, \textit{connection establishment}, where the actual connection between the client and the proxy is created, and \textit{data transfer}, where the proxy relays communication between the client and a Tor bridge. Our focus in this paper is mainly on the rendezvous and connection establishment phases. In the rendezvous phase, the client sends a request to the centralized Snowflake broker through a highly censorship-resistant signaling channel (\eg Domain Fronting)~\cite{fifield2015blocking, pu2024exploring}. The broker subsequently matches the client's request to a proxy that has previously \textit{polled} its availability before the two begin communicating with each other directly using the WebRTC protocol.

To facilitate the connection establishment, the client and proxy attempt to determine their NAT type through a series of STUN requests~\cite{rfc5780}, or a WebRTC connection to a probe server behind a restrictive NAT, respectively. The NAT types of the client and proxy play a large role in the matching performed by the broker, as some NAT mapping and filtering types are not compatible. Under the current Snowflake design, proxies that successfully establish a peer-to-peer connection with a probe server behind a symmetric NAT are referred to as \textit{unrestricted} and are matched with clients that have a self-reported restricted NAT type. Unrestricted proxies are typically high-capacity and run as \texttt{standalone} programs on VPS or cloud hosts. \textit{Restricted} or \textit{Unknown} proxies, matched with clients that report an unrestricted NAT type, are more plentiful and commonly run as extensions on volunteer browsers or as mobile apps, such as Orbot's Kindness Mode~\cite{orbotkindness}.

The broker's design is central to both our attack model and our analysis. Snowflake uses a centralized, client-accessible broker that maintains available proxy sessions and matches proxies to clients over public HTTP interfaces~\cite{snowflake-code, bocovich2024snowflake}. When a client contacts the broker, it sends an SDP offer; proxies simultaneously poll the broker, advertising metadata (NAT type, self-reported load), and the broker prioritizes lower-load proxies. If no client is waiting, the broker replies “no clients” after a timeout (currently 10 s) and the proxy idles until its next poll; otherwise, it forwards the client’s SDP offer to the proxy, which returns an SDP answer that the broker relays to the client. Thereafter, client and proxy communicate directly, with the broker serving only as a WebRTC signaling server.

\section{Threat Model}
\label{sec:threat-model}
We evaluate two linked \textit{practical} attacks on proxy-based censorship circumvention systems as shown in Figure~\ref{fig:attack-overview}: (1) \textit{Enumeration:} An adversary acts as a client in the communication system and enumerates the proxy IP addresses that are returned by the broker. By doing so, an adversary can gain knowledge of proxy concentrations across locations and networks, violating privacy while gaining targets to block. (2) \textit{Blocking:} Based on these enumeration efforts, adversaries may choose to block certain proxy IP addresses or entire networks with high proxy concentrations. Successful blocking attacks may render proxy-based communication systems unusable for circumvention.

For our enumeration attack, we model our attackers as active malicious clients within the Snowflake ecosystem. Our attacker does not require privileged access, large-scale infrastructure, or control over the network. They can be instantiated using ordinary clients that repeatedly interact with the public broker interface and record returned proxy IP addresses. While we bound our real-world measurements due to safety considerations (refer Appendix~\ref{sec:ethics}), we note that Snowflake currently does not impose any rate limits on client polls. As a result, a practical real-world adversary can deploy multiple clients in parallel, increasing enumeration coverage as we show in our simulation experiments.

For our blocking attack, we assume the adversary has the ability to block access to either individual proxy IP addresses (that result from enumeration, for example) or entire networks (at the AS level). These capabilities are practical for nation-state adversaries~\cite{ramesh2019decentralized,cui2026characterizing}.  Although real-world adversaries typically affect only clients within their jurisdiction, our simulations assume that blocked proxies become inaccessible to all clients for clarity.

Both of the above attacks place the adversary in the client role. Because Snowflake imposes no rate limits on the proxy role either, we also consider a case study of an adversary that registers as a malicious proxy to influence its own position in the broker's heap, targeting clients rather than proxies (\S\ref{sec:malicious-proxy}).

\section{Methods and Experiments}
\label{sec:method}
To evaluate Snowflake's resistance to enumeration and blocking attacks, we combine real-world measurements with controlled simulation. Real-world measurements are necessary to capture the behavior of the live broker, including proxy churn, pool composition, and demand-driven effects that arise in deployment. At the same time, extensive measurements against a live circumvention system raise important ethical concerns, which we discuss in detail in Appendix~\ref{sec:ethics}. Controlled simulations therefore let us study counterfactuals, vary parameters that cannot be safely manipulated in production, and evaluate defenses without imposing additional burden on a system that serves real users. We thus use the two methods in tandem, with the simulation design grounded in observations from our real-world measurements.

\myparagraph{Summary of Ethics Considerations.} Our study involved detailed deliberations of the safety of our experiments given experiments against a live circumvention system. We provide a detailed discussion of our ethics considerations in Appendix~\ref{sec:ethics}, and provide a summary of key principles here to guide our experiment design: 
\begin{wideitemize}
\item \textbf{Independent review and coordination.} We developed the study design through formal review by the Tor Research Safety Board~\cite{tor-safety-research-board} and repeated consultation with Snowflake developers, and grounded our process on the Menlo Report and prior Tor research~\cite{menlo, jansen2021once,bauer2011experimentor,jansen2011shadow}.
\item \textbf{Minimizing operational harm.} We performed aggressive evaluation of attacks only in simulation, limiting live measurements to only two bounded, low-rate probers. We also never completed connections to enumerated proxies, so that they remained available to legitimate users.
\item \textbf{Protecting proxy operators.} We did not store raw proxy IP addresses on disk. Instead, we stored only ephemeral keyed hashes for duplicate detection and released only aggregated data, reducing the risk of exposing volunteers. 
\item \textbf{Adapting to real-world risk.} When demand for Snowflake proxies rose sharply during the June 2025 Iran event, we ended the live measurements early rather than continue probing during a period of heightened operational need.
\end{wideitemize}

\subsection{Real-world Measurements}
\label{sec:real-world-measurements}
Our real-world measurements use a modified Snowflake client to emulate a practical adversary. The malicious client polls the live Snowflake broker as an ordinary client, records only the minimum information required for later analysis, and terminates after receiving a broker response.
We use two separate probers because a single client configuration cannot observe both NAT-specific proxy pools in Snowflake. One prober targets the larger \textit{restricted pool} (with an average of 159,700 proxies over all days in June 2025~\cite{snowflake-broker-metrics}) by advertising an unrestricted client NAT, and waits at least 1 second between requests. The second prober targets the smaller \textit{unrestricted pool} (with an average of 2,400 proxies over all days in June 2025~\cite{snowflake-broker-metrics}) by advertising a restricted client NAT and waits at least 10 seconds between requests.

Each successful broker response reveals a proxy IP address to the client, which we map to its origin AS using RouteViews prefix-to-AS data~\cite{routeviews-prefix2as}. We then compute a keyed SHA256 hash of the IP address for duplicate detection, with the key generated only in memory and discarded on exit. We only store the hash and its corresponding AS number on disk, \ie, the stored dataset never contains raw proxy IP addresses. Although this design decision prevents us from resuming enumeration after stoppage, it protects against accidental leaks.

We designed the live collection as a longitudinal measurement, and our analysis covers 48 days from May~8, 2025 through June~24, 2025. When client demand in the deployed Snowflake ecosystem increased due to the shutdown in Iran in mid-to-late June 2025, we stopped our experiments to avoid increasing load on the system~\cite{cui2026characterizing,snowflake-june-2025,net4people-iran-june-2025,snowflake-iran-june-2025}.  Over the May~8 to June~24 analysis window, the restricted-pool prober recorded 399{,}118 broker responses yielding 13{,}152 unique proxy IP hashes spanning 935 ASNs, while the unrestricted-pool prober recorded 159{,}086 broker responses yielding 8{,}291 unique proxy IP hashes spanning 810 ASNs. %

We enrich the resulting dataset with additional public metadata to support the blocking attack analysis. We use data from IPInfo to determine the type of AS network: IPInfo classified ASes as ``ISP'' (typically access or residential networks), ``Hosting'', ``Business'', ``Government'', and ``Education'' ~\cite{ipinfo}. We also obtain the list of Top~1M popular domains from Tranco~\cite{tranco} and resolve them to estimate the service-side context needed to estimate collateral damage when blocking networks. Finally, we use the AS population dataset from APNIC to estimate the number of users in ASes~\cite{stats-apnic,salamatian2024s} and use CAIDA's AS Rank information to estimate AS cone sizes~\cite{as-rank}. 

\subsection{Simulation Experiments}
\label{sec:simulation-experiment}
\begin{table*}[t]
\centering
\small
\begin{tabular}{p{2cm}p{3.5cm}p{4cm}p{7cm}}
\toprule
\textbf{Parameter} & \textbf{Default Value} & \textbf{Experimental Values} & \textbf{Experiment Description} \\
\midrule
Attacker Fraction & 0.05\% of clients &
\texttt{0.001\%}, \texttt{0.01\%}, \texttt{0.1\%}, \texttt{0.5\%} &
Change attacker population to 0.001\%, 0.01\%, 0.1\%, and 0.5\% of the total number of clients. \\
\midrule
Churn Rate (Uniform) & 2.7\% hourly &
\texttt{0.25x}, \texttt{0.5x}, \texttt{2x}, \texttt{4x} & Change hourly proxy churn rates to 0.675\%, 1.3\%, 5.4\%, 10.8\%. \\
\midrule
Standalone Churn Rate & 2.7\% hourly & 1.35\%, 0.27\% & Reduce only standalone-proxy churn to one-half and one-tenth of the default rate while holding WebExt and IPT churn constant (\S\ref{sec:standalone-case-study}). \\
\midrule
Connection Duration & 3 hours $\pm$ 30 minutes & 
 \texttt{30m$\pm$5m}, \texttt{1.5h$\pm$15m}, \texttt{6h$\pm$1h}, \texttt{9h$\pm$1.5h} & 
Change client-proxy connection duration to means of 30m, 1.5h, 6h, and 9h and standard deviation of 5m, 15m, 1h and 1.5h respectively. \\
\midrule
\bottomrule
\midrule
\rowcolor{gray!30}
Malicious Proxy Attack & Default polling rate with self-reported client load &
\texttt{1s} polling rate with fake low client-load ($-1$ clients) &
A practical attack where an adversary operates a malicious proxy that polls aggressively and self-reports a fake client load to consume a majority of offers. (\S\ref{sec:malicious-proxy}). \\
\bottomrule
\midrule
\rowcolor{gray!30}
Equal Polling Defense & \texttt{standalone}: 5s, \texttt{webext}: 60s, \texttt{iptproxy}: 120s &
\texttt{10s}, \texttt{120s}, \texttt{240s} &
A potential mitigation where the broker instructs \textit{all} proxies to poll at intervals of 10, 120 or 240 seconds (\S\ref{sec:mitigations}). \\
\bottomrule
\end{tabular}
\caption{The various simulation parameters we evaluate in our experiments, along with their default values. We also evaluate a practical malicious proxy attack as a case study in \S\ref{sec:malicious-proxy} and highlight settings for a potential mitigation strategy we discuss in \S\ref{sec:mitigations}.}
\label{tab:different-settings}
\end{table*}

Because our real-world measurements are deliberately bounded for safety, they cannot evaluate stronger attackers or defenses under controlled conditions. We therefore complement the live study with a simulation framework that lets us vary attacker scale, proxy churn, connection duration, and defense settings while preserving the broker behavior most relevant to enumeration and blocking.

Our simulation design is focused on two goals: (1) performing simulations at 100\% network scale, as enumeration and blocking attacks are significantly affected by the scale of the system under test~\cite{jansen2021once}, and (2) evaluating attacks longitudinally (30-days) as this shows the long-term effect of proxy churn. However, designing simulations that satisfy both these constraints faces several implementation challenges. During our initial design, we considered both adapting the Shadow simulator~\cite{jansen2021once,jansen2011shadow,tracey2018high} and building a process-level emulation using Snowflake's open-source client, proxy, and broker code~\cite{snowflake-code}. In both cases, the resource cost of maintaining full-scale network state across tens of thousands of proxies and clients was too high, forcing scaled-down networks.

To resolve these challenges, we build a Snowflake simulation tool from scratch that combines existing Snowflake code with lightweight abstractions of Snowflake client and proxy processes to enable large-scale experiments. The simulator preserves the part of Snowflake that matters most for our attacks: the broker's view of available proxies and its resulting matching decisions. We directly run the Snowflake broker code in a dedicated simulation mode and reuse its IPC interfaces, NAT-aware matching logic, proxy selection, and timeout handling, but do not simulate full packet-level TCP, \mbox{WebRTC} or Tor traffic.  Clients, proxies, and attackers are modeled as lightweight stateful threads that exchange synthetic offers and polls sufficient to exercise the broker path. This design scales to 100\% network size, with one simulation step corresponding to one fake second, and remains deterministic while preserving the asynchronous broker interactions that determine attacker visibility. 

 To support repeated 30-day experiments at the scale of the deployed Snowflake network, our simulator abstracts away the data-plane communication that follows broker matching, including full Tor traffic and \mbox{WebRTC}, network latency, packet loss, and transport failures. These factors can affect connection timing and the number of retries experienced in deployment, but their effects on attack outcomes cannot be modeled reliably. We therefore use the simulator to compare attack outcomes across clean, controlled settings rather than to predict exact deployment-level connection times or failure rates. We focus on comparisons across experimental settings and treat absolute coverage and retry values from the simulations as model-dependent estimates.

We implement our simulation in GoLang. Our lightweight simulation framework is able to perform a 30-day simulation in about 2 real days on an Ubuntu server with 12 vCPUs and 70GB memory.
All simulation results in \S\ref{sec:simulation-results} are computed over three independent simulation runs, and we report the mean and standard deviation. Because the simulator is highly deterministic, we observe only little variation across runs.
We open-source our simulation so that it can continue to be used to evaluate Snowflake under various adversarial settings (See Appendix~\ref{sec:open-science}). 

A central design challenge in our simulations is that Snowflake's public metrics expose only some of the quantities required by the simulator, so certain parameters must be modeled rather than directly inferred from the metrics~\cite{snowflake-broker-metrics, snowflakeprometheus}. We describe each of our simulation experiment settings in detail below and in Table~\ref{tab:different-settings}.

\subsubsection{Default Simulation Settings}
\label{sec:default-settings}
We first describe the settings we implement for our default baseline simulation. 

\myparagraph{Snowflake Proxies.} Proxy populations are driven by 30 days of Snowflake metrics for the three main proxy classes: \texttt{standalone}, \texttt{webext}, and \texttt{iptproxy}. \texttt{Standalone} proxies are typically high-capacity, unrestricted-NAT machines that poll every 5 seconds and can serve many clients~\cite{bocovich2024snowflake}. In contrast, \texttt{webext} and \texttt{iptproxy} proxies are volunteer-run browser and mobile proxies that poll every 60 and 120 seconds, respectively, and typically serve only one client at a time when behind restrictive NATs~\cite{snowflake-web-code,orbot-android-code,orbot-apple-code}.
Using public Snowflake metrics, the simulator stores 720 hourly target values for each proxy type and applies them at each simulated hour. Across these traces, \texttt{standalone} proxies range from 3{,}636--6{,}091 (mean 4{,}766), \texttt{webext} from 46{,}452--115{,}135 (mean 77{,}664), and \texttt{iptproxy} from 30{,}541--67{,}964 (mean 42{,}350). To avoid artificial hour-boundary spikes, target-count changes are spread randomly across the hour. \texttt{Standalone} proxies continue polling after successful matches to model their higher serving capacity, whereas \texttt{webext} and \texttt{iptproxy} proxies stop polling while assigned to a client and return only when the simulated connection ends. Initial proxy poll times are spread across the first 24 simulated hours.

\myparagraph{Proxy Churn.} We set the default base proxy churn rate as 2.7\% per hour from the results reported in the original Snowflake study~\cite{bocovich2024snowflake}. This constant-rate model keeps the simulations tractable and isolates the comparative effect of churn without overfitting to real-world observations, but does not reproduce non-memoryless proxy lifetimes or temporal and geographic variation in deployment, which we concede as a deliberate limitation. Because standalone proxies are likely to be more stable than browser and mobile proxies, we separately perform a case study in \S\ref{sec:standalone-case-study} of standalone proxies churning at one-half (1.35\% per hour) and one-tenth (0.27\% per hour) of the default churn rate.

\myparagraph{Snowflake NAT Compatibility.} In the default setting, \texttt{standalone} proxies are unrestricted with probability 0.9, while \texttt{iptproxy} and \texttt{webext} proxies are unrestricted with probability 0.1, based on Snowflake metrics~\cite{snowflakeprometheus}. We empirically observe that alternate NAT distributions do not result in significant changes in attack success in our simulations. 

\myparagraph{Clients.}  Based on Snowflake’s published metrics~\cite{snowflake-clients}, we simulate roughly 61{,}600 clients regenerated hourly throughout the experiment. Client first-poll times are spread over the first 24 simulated hours, and client NAT type is sampled on each poll with a probability 0.72 for unrestricted and 0.28 for restricted according to public metrics~\cite{snowflakeprometheus}. Clients that fail to obtain a usable proxy retry after 3 simulated seconds. 

\myparagraph{Client-Proxy Connections.} This setting determines how long \texttt{webext} and \texttt{iptproxy} proxies remain unavailable after a successful match with a client. In the default setting, we sample the connection duration from a truncated normal distribution with mean of 3 hours and standard deviation of 30 minutes. %

\myparagraph{Attackers.} Attackers are modeled as malicious clients. Unlike in our real-world study, the default simulation uses multiple attacker probers: 0.05\% of the total client pool is converted into malicious clients (32 attackers). After a 24-hour warm-up period, each attacker polls once per simulated second, with attackers split evenly by NAT type to enumerate both proxy pools. In \textit{enumeration-only} mode, the simulator records each first-seen proxy ID. In \textit{blocking} mode, each enumerated proxy is immediately added to a blocklist, so later benign clients matched to blocked proxies must retry.

\subsubsection{Simulation Experiments.}
\label{sec:experiments-list}
Next, we describe the key simulation parameters we vary to evaluate enumeration and blocking under different system conditions (see Table~\ref{tab:different-settings}):

 \myparagraph{Attacker Fraction Experiment.}  We evaluate attacker fractions of 0.001\%, 0.01\%, 0.1\%, and 0.5\%, in addition to the default 0.05\% malicious clients. These experiments measure the impact of attacker scale on the effectiveness of enumeration and blocking attacks. Since Snowflake does not enforce rate limits, increasing attacker count is equivalent to increasing aggregate poll rate.

 \myparagraph{Churn Rate Experiment.}  Proxy churn is Snowflake's primary defense against enumeration and blocking attacks.  We vary the default proxy churn of 2.7\% per hour across four additional settings: 0.675\%, 1.3\%, 5.4\%, and 10.8\% per hour. These experiments measure how changes in proxy populations affect our attacks, and can guide future proxy recruitment efforts. 

  \myparagraph{Connection Duration Experiment.} The duration that a matched proxy remains unavailable after matching with a client affects both discovery and retry behavior.  We evaluate two additional duration distributions shorter (1.5h$\pm$15m, 30m$\pm$5m) and two longer (6h$\pm$30m, 9h$\pm$1.5h) than the default.

\section{Results}
\label{sec:results}
We first discuss the results from our real-world measurements (\S\ref{sec:real-world-results}) before discussing the effect of various simulation settings (\S\ref{sec:simulation-results}). 

\begin{figure*}[!t]
  \includegraphics[width=0.99\linewidth]{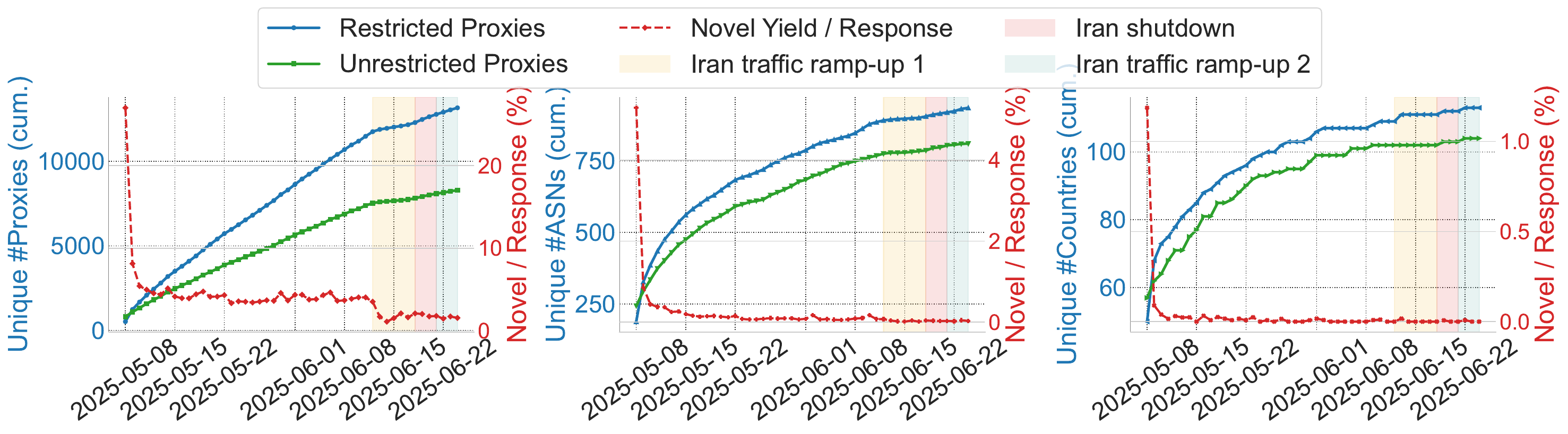}
  \caption{Cumulative Enumeration Counts of proxies in Real-World Experiments. The attacker rapidly accumulates unique restricted and unrestricted proxy IP hashes, ASNs, and countries (left Y-axis), but the percentage of new proxies found per successful response (Novel / Response, right Y-axis in red) declines over time due to proxy churn.}
  \label{fig:enumeration-proxies-asns-countries}
\end{figure*}

\begin{figure}[!t]
  \includegraphics[width=0.95\linewidth]{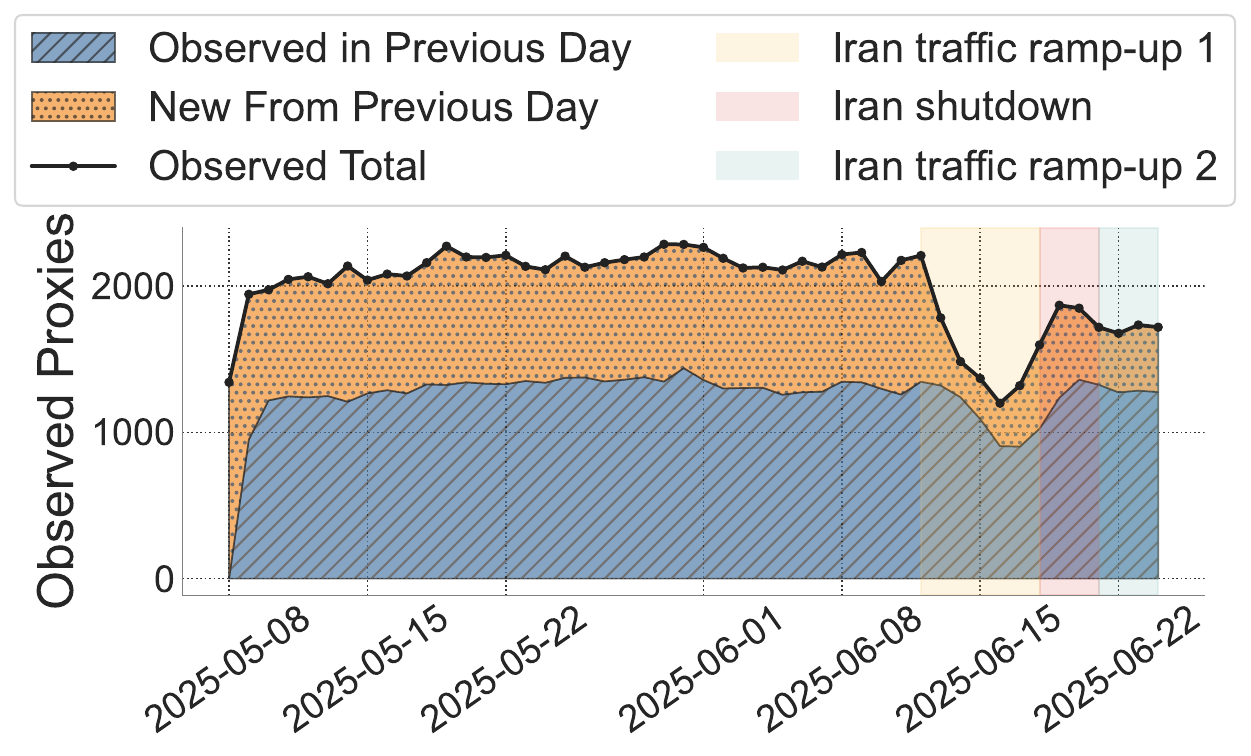}
  \caption{Daily observed and retained proxies. During the Iran traffic ramp-up and shutdown period, the number of active proxy IPs visible to the attacker drops sharply.}
  \label{fig:active-retained-proxies-asns}
\end{figure}

\subsection{Real-World Experiments}
\label{sec:real-world-results}
\subsubsection{Proxy Enumeration.} 
Our real-world proxy enumeration attack (\S\ref{sec:real-world-measurements}) rapidly discovers a large and diverse set of proxies, but the rate of discovery slows and stabilizes over time. Figure~\ref{fig:enumeration-proxies-asns-countries} shows that, over our 48-day measurement window (May~8--June~24, 2025), our attackers enumerated 13{,}152 unique restricted proxy IPs and 8{,}291 unique unrestricted proxy IPs, across a total of 935 and 810 ASNs and 113 and 104 countries, respectively.
Relative to Snowflake’s public broker metrics, which report an average daily pool of $\approx$150{,}000 restricted IPs and $\approx$2{,}500 unrestricted IPs, our attacker's coverage is especially substantial for the unrestricted pool, but remains bounded by our slow probing rate and, more fundamentally, by proxy churn. Churn also explains the attack’s front-loaded success: our attackers reached 50\% of their final observed proxy IP tallies by May~26, and 80\% by June~8, while continued probing yields stable returns. The number of newly discovered IPs per 1{,}000 successful responses falls from 67.3 in the first week to 17.7 in the final week. Together, these results show that proxy churn does not prevent substantial early enumeration, but it does limit how completely a practical attacker can sustain visibility into the Snowflake ecosystem over time, and is consistent with Snowflake's lightweight design~\cite{bocovich2024snowflake}.

However, the observed enumeration is not a uniform sample of the underlying Snowflake proxy population. Instead, the broker's NAT-aware matching and proxy's current capacity- and load-based prioritization preferentially expose stable, higher-capacity proxies to attackers. The top 1\% of returned proxies account for 35.8\% of observed restricted and 28.3\% of observed unrestricted pools, and the most frequently returned proxy appears 15{,}830 and 6{,}663 times, respectively. These repeatedly surfaced IPs remain stable throughout our measurement period, suggesting that they are likely large standalone proxies with high client-serving capacity. Thus, although a long tail of short-lived proxies remains difficult to enumerate fully, the attacker quickly learns the small subset of proxies that contributes disproportionately to both observed enumeration and likely blocking impact, since these same high-capacity proxies are also more likely to serve benign clients.

\takeaway{Proxy churn prevents complete and sustained enumeration of proxy IP addresses, but even modest probing quickly reveals a stable, high-capacity subset of proxies that are disproportionately important for later blocking.}

\begin{table}[!t]
\centering
\footnotesize
\caption{Number of Restricted (R) and Unrestricted (U) Proxies (\#P) and ASNs (\#A) in each AS Type. We observe that more than 80\% of proxies are in ISP (access) networks. We also show the average number of proxies and ASNs observed per day during the initial period and during the shutdown in Iran, when ISP proxies decreased due to high demand.}
\label{tab:as-type-ip}
\begin{tabular}{p{1cm} >{\raggedleft\arraybackslash}p{1.3cm} >{\raggedleft\arraybackslash}p{1.2cm} >{\raggedleft\arraybackslash}p{1.6cm} >{\raggedleft\arraybackslash}p{1.55cm}}
\toprule
\multirow{2}{*}{\textbf{AS Type}} &
\multicolumn{2}{c}{\textbf{Full Period}} &
\textbf{Avg./Day} &
\textbf{Avg./Day} \\
\cmidrule{2-3} \cmidrule(lr){4-5} & \textbf{R} & \textbf{U}
& \textbf{Pre-Iran}
& \textbf{Iran $\uparrow$} \\
& \textbf{\#P (\#A)} & \textbf{\#P (\#A)}
& \textbf{(to Jun 11)}
& \textbf{(Jun 12--17)} \\
\midrule
ISP        & 11,033 (642) & 6,750 (543)  & 1033 (209) & 552 (129) \\
Hosting    & 1,557 (216)  & 1,245 (204)  & 1006 (124) & 951 (120) \\
Education  & 202 (38)     & 100 (32)         & 24 (9)     & 16 (6)    \\
Business   & 99 (28)      & 49 (21)         & 12.0 (7)     & 7.8 (4)     \\
Unknown    & 174 (11)     & 100 (8)          & 41.4 (5)     & 31.8 (4)    \\
Govt. & 87 (1)       & 47 (3)    & 5.0 (1)      & 2.2 (1)     \\
\midrule
\textbf{Total} & \textbf{13,152 (935)} & \textbf{8,291 (810)} & \textbf{2121 (354)} & \textbf{1561 (265)} \\
\bottomrule
\end{tabular}
\end{table}

\myparagraph{Impact of Increasing Demand.} The Iran-related surge in June 2025 changed the composition of the attacker-observable proxies. Figure~\ref{fig:active-retained-proxies-asns} shows that active IPs across both proxy pools fell from 2{,}210 on June~12 to 1{,}369 on June~15, a 38.1\% decrease, while active ASNs fell from 360 to 231, a 35.8\% decrease.
Table~\ref{tab:as-type-ip} shows that this was accompanied by a clear difference in the type of proxy networks that the attacker was able to enumerate. Before the Iran ramp-up period (May~8--June~11), the attacker was able to observe 1{,}033 proxies/day on average in \textit{ISP} networks, but during \texttt{Iran Ramp-Up~1} (June~12--June~17) they fell to 552 proxies per day, a decrease of 46.5\%. Proxies in \textit{Hosting} networks on the other hand were far more stable, falling only from 1{,}006 to 951 proxies/day (5.5\% $\downarrow$). As corroborated by reports by Snowflake~\cite{snowflake-iran-june-2025,snowflake-june-2025}, the number of clients in Iran increased rapidly during the \texttt{Iran Ramp-Up~1} phase, likely utilizing a large majority of lightweight ISP proxies, which can only handle one client and are hence removed from the proxy pool while in use. Therefore, the attacker was able to only receive the large \texttt{standalone} proxies primarily hosted in Hosting networks, which can handle multiple clients. Overall, our experiment shows that although Snowflake can absorb periods of increased demand, these periods also expose attacker-visible shifts in the proxy population, enabling an adversary to further identify and block large \texttt{standalone} proxies and thereby place even greater load on lightweight \texttt{webext} and \texttt{ipt} proxies.

\takeaway{The June 2025 Iran event shifted the attacker’s view toward a smaller but more stable, infrastructure-heavy subset of proxies with large capacity.}

\begin{table}[!t]
\centering
\footnotesize
\caption{Top Countries and ASNs by Proxy Count. DE, US and IN have the most observed proxy IPs.}
\label{tab:top-countries}
\begin{tabular}{p{0.2cm} >{\raggedleft\arraybackslash}p{1cm} p{6.2cm}}
\toprule
\textbf{CC} & \textbf{\# Proxies} & \textbf{Top ASes in CC (\# proxies)} \\
\midrule
DE & 10,134 & VODANET (AS3209, 4,075); TDDE-ASN1 (AS6805, 1,449); VERSATEL (AS8881, 1,375) \\
US & 1,744  & AMAZON-02 (AS16509, 200); COMCAST-7922 (AS7922, 186); T-MOBILE-AS21928 (AS21928, 164) \\
IN & 1,057  & RELIANCEJIO-IN (AS55836, 603); BHARTI-MOBILITY-AS-AP (AS45609, 230); AIRTELBROADBAND-AS-AP (AS24560, 52) \\
\bottomrule
\end{tabular}
\end{table}

\myparagraph{Network Characteristics of Enumerated proxies.}
The enumerated proxies are geographically diverse but structurally concentrated across networks and countries. Table~\ref{tab:top-countries} shows that although our attacker eventually observed proxies in more than 100 countries, the distribution is heavily skewed: Germany alone accounts for 10{,}134 (47.3\%) unique proxies, followed by the United States with 1{,}744 (8.1\%), India with 1{,}057 (4.9\%), Iran with 871 (4.1\%), and France with 720 (3.4\%). Together, these top five countries account for 14{,}526 of 21{,}443 combined unique proxy IP addresses (67.7\% of all enumerated proxies). The concentration is also visible within countries: in Germany, the top three ASNs alone---AS3209 (4{,}075 proxies), AS6805 (1{,}449), and AS8881 (1{,}375)---account for 68.1\% of the proxies in Germany.

Table~\ref{tab:as-type-ip} shows that proxies are also concentrated by network type: 17{,}783 of 21{,}443 enumerated proxies (82.9\%) are hosted in ISP networks, which are typically residential access networks, while Hosting providers account for another 2{,}802 (13.1\%). As we show later, this concentration makes network-level blocking effective while still imposing relatively little collateral cost.
This concentration is a property of not only the underlying deployment, but also of the attacker's observation process. Because the broker's matching and load balancing preferentially exposes stable, higher-capacity proxies, enumeration over-represents networks that contribute disproportionate connectivity, reinforcing the prominence of infrastructure-heavy ASes in attacker-visible data.

\takeaway{A small number of countries, ASes, and AS types account for a large fraction of the enumerated proxies.}

\begin{figure}[!t]
  \includegraphics[width=0.99\linewidth]{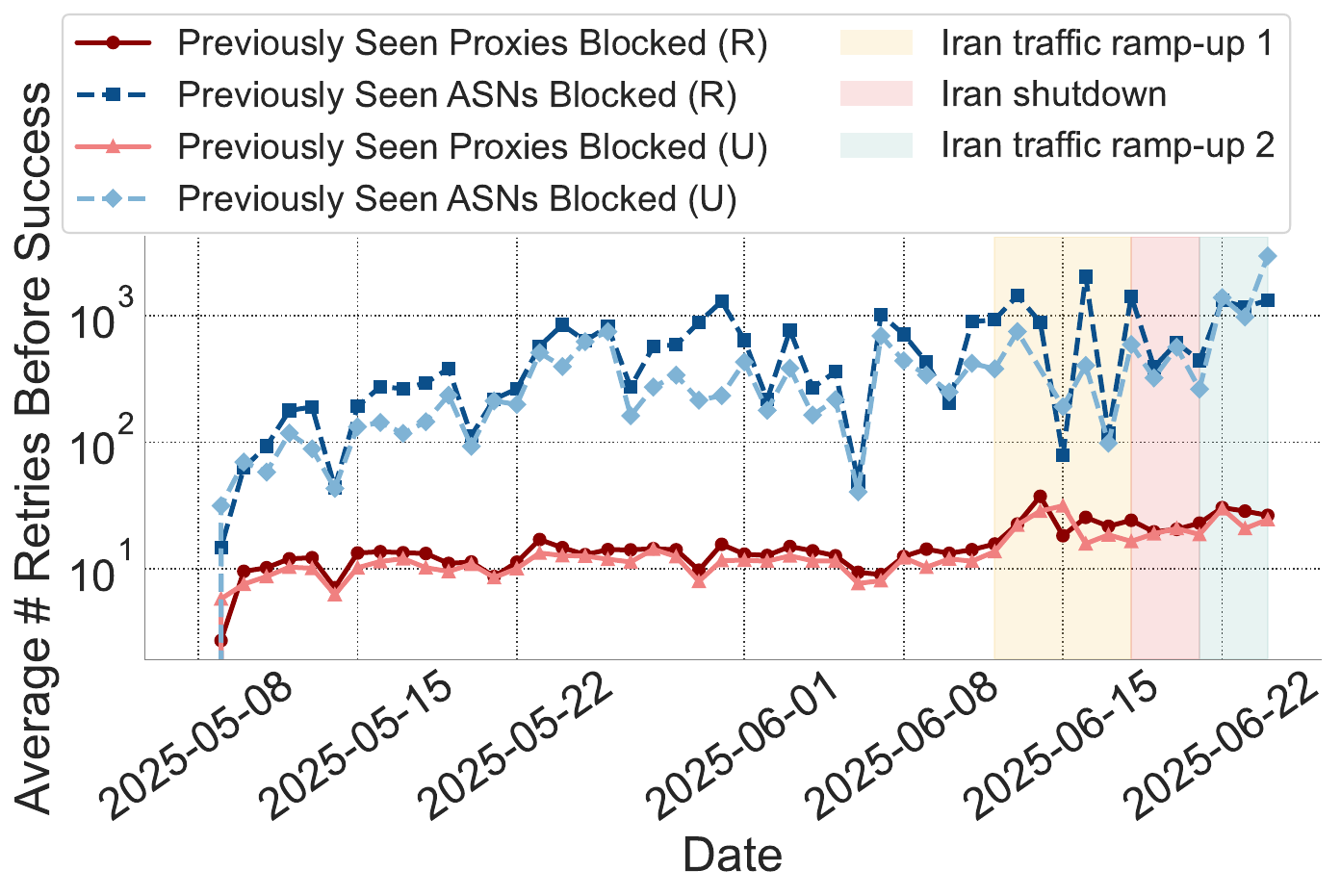}
  \caption{Estimated retry burden on each day under IP and ASN-level blocking (Note: Y-axis is in log scale). Blocking proxy IPs seen on previous days only modestly increases expected retries, while blocking ASNs raises expected retries into the hundreds and thousands.}
  \label{fig:blocking-retries-previously-seen}
\end{figure}

\subsubsection{Proxy Blocking.} 
\label{sec:real-world-blocking}
We next evaluate the real-world blocking attack by approximating the effect of an attacker blocking proxy IP addresses or networks immediately after enumerating them. Because the attacker is itself a client and only learns a proxy when the broker returns it, we model the attacker's blocklist on each day $d$ as the set of unique proxy IPs or ASNs observed before that day ($<d$). We then estimate the retry burden for clients on $d$ as: if a fraction $p$ of responses return a proxy or ASN not already on the blocklist, the expected number of retries before success is $(1-p)/p$. This gives an average retry-cost estimate for IP-level and ASN-level blocking under the assumption that benign clients see the same broker-visible distribution as our probers.

 We see from Figure~\ref{fig:blocking-retries-previously-seen} that attackers blocking proxies enumerated on previous days substantially increases expected client retry cost, but IP churn keeps the number of retries stable over time. The average expected retries rise to 26.5 for restricted proxies and 24.6 for unrestricted proxies by June~24. At the same time, the blocklist does not remain perfectly current because new IPs continue to appear: we perform a blocklist-stability analysis (see Appendix~\ref{sec:app-real-world-blocklist-stability}) to observe that an IP blocklist frozen on the day $d$ still covers 79.3\% of the next day’s observed unique proxies, but this falls to $\sim$65\% 2 weeks after $d$. This shows that a censor must continuously probe and refresh their blocklist, rather than enumerate once and block forever. Compared with Kon \etal~\cite{kon2024spotproxy} and Fares \etal~\cite{fares2026game}, who do not quantify the decay of a real blocklist, our measurements show that IP blocking is operationally costly because churn remains active over time. 
 
\myparagraph{Blocking Entire Networks.} Blocking ASes enumerated in previous days is dramatically more powerful than blocking individual IP addresses. Figure~\ref{fig:blocking-retries-previously-seen} shows that under ASN-level blocking (where every IP in ASNs enumerated on previous days is blocked), the expected retry count rises to 80 for restricted proxies and 194 for unrestricted proxies by June~15, and reaches 1{,}349 and 3{,}014 by June~24, respectively. Figure~\ref{fig:blocking-frequency} in Appendix~\ref{sec:app-real-world-blocking-strategies} helps explain these significant retry values: the single most frequent AS already covers $\approx$30\% of all observed proxies, the top 5\% of observed ASes cover more than 50\% of observed proxies.  Thus, although proxy churn limits the value of IP blocking, the concentrated network-level structure of Snowflake still gives an attacker a very strong network blocking primitive. Our real-world experiments collecting network information show, for the first time, how powerful network-level blocking can be against proxy-based circumvention tools. 

\takeaway{IP-level blocking is disruptive but manageable for Snowflake clients currently due to high proxy churn. However, AS-level blocking is the real threat in deployment, driving retry burden for clients into the thousands.}

\begin{figure}[!t]
  \includegraphics[width=0.95\linewidth]{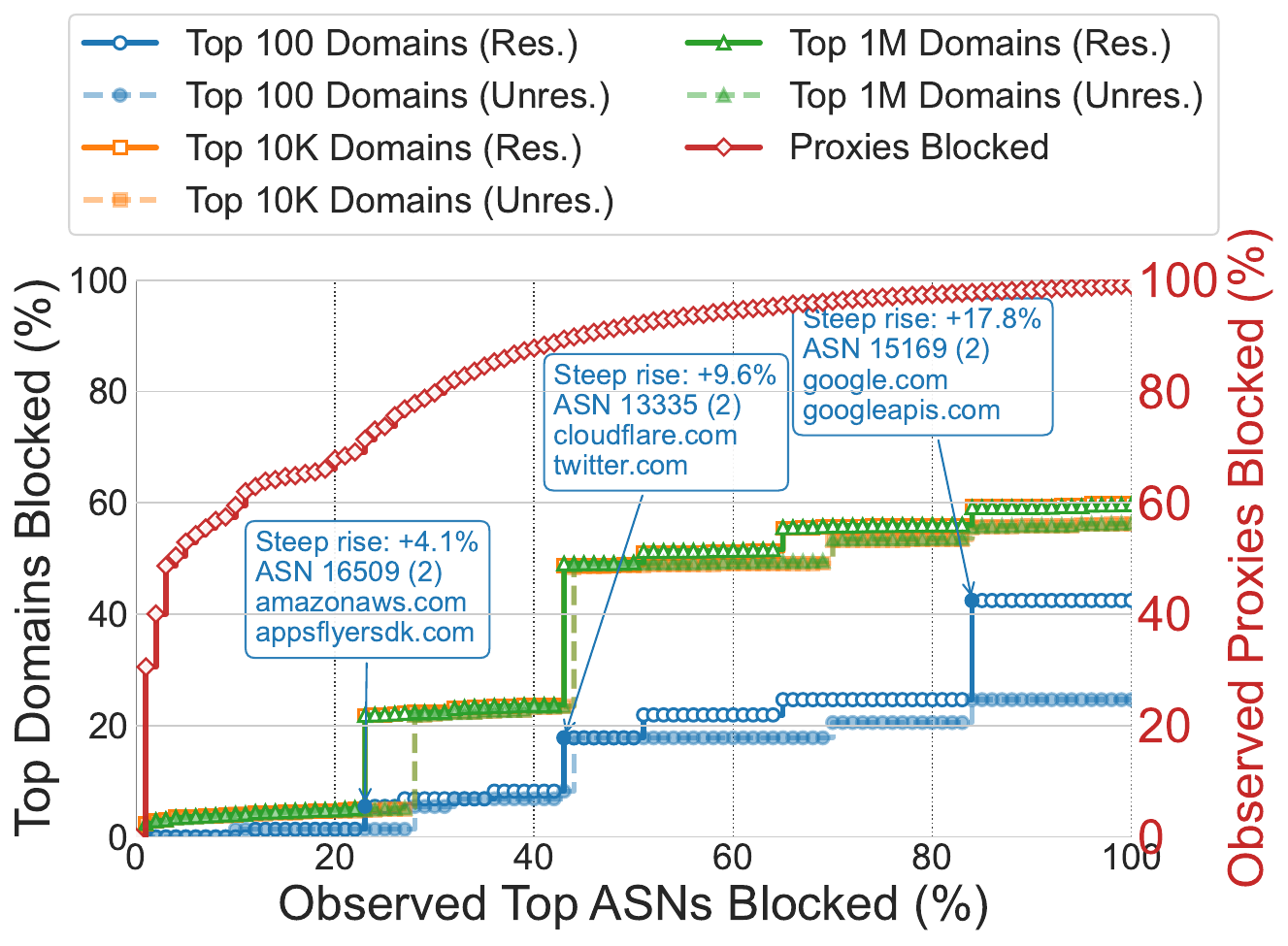}
  \caption{\textbf{Collateral damage from blocking increasingly frequent proxy ASes.} Attackers can block >30\% of proxies by blocking just the top 1\% of frequently observed ASNs, while affecting 0\% of Top~100 domains.}
  \label{fig:blocking-collateral-domains}
\end{figure}

\myparagraph{Collateral Damage of Network-Level Blocking.} A natural barrier against network-level blocking is the high collateral damage caused by inadvertent blocking of important services such as popular websites hosted in the same network. We next investigate the collateral cost of blocking frequently appearing ASes in our real-world measurements.

Figure~\ref{fig:blocking-collateral-domains} shows that the censor can block a large fraction of proxies while affecting relatively few popular domains: blocking the top 1\% of most frequent ASes already covers 30.1\% of restricted proxies and 30.9\% of unrestricted proxies, while blocking 0\% of the Tranco Top~100 domains, 1.55\% of the Top~10K domains, and 2.45\% of the Top~1M domains. At 5\% of observed ASes, proxy coverage rises to over 50\%, while Top~100 collateral remains 0\% and Top~1M collateral remains only $\approx$3.5\%. Once major infrastructure ASes enter the blocklist, collateral rises abruptly. For example, adding ASN~16509 (\texttt{AMAZON-02}) raises Top~100 collateral from 1.37\% to 5.48\%, and adding ASN~13335 (\texttt{CLOUDFLARENET-AS}) raises it from 8.22\% to 17.81\%.  Our findings suggest that the censor’s best strategy must involve knowledge of high-collateral networks.

\begin{figure}[!t]
  \includegraphics[width=0.99\linewidth]{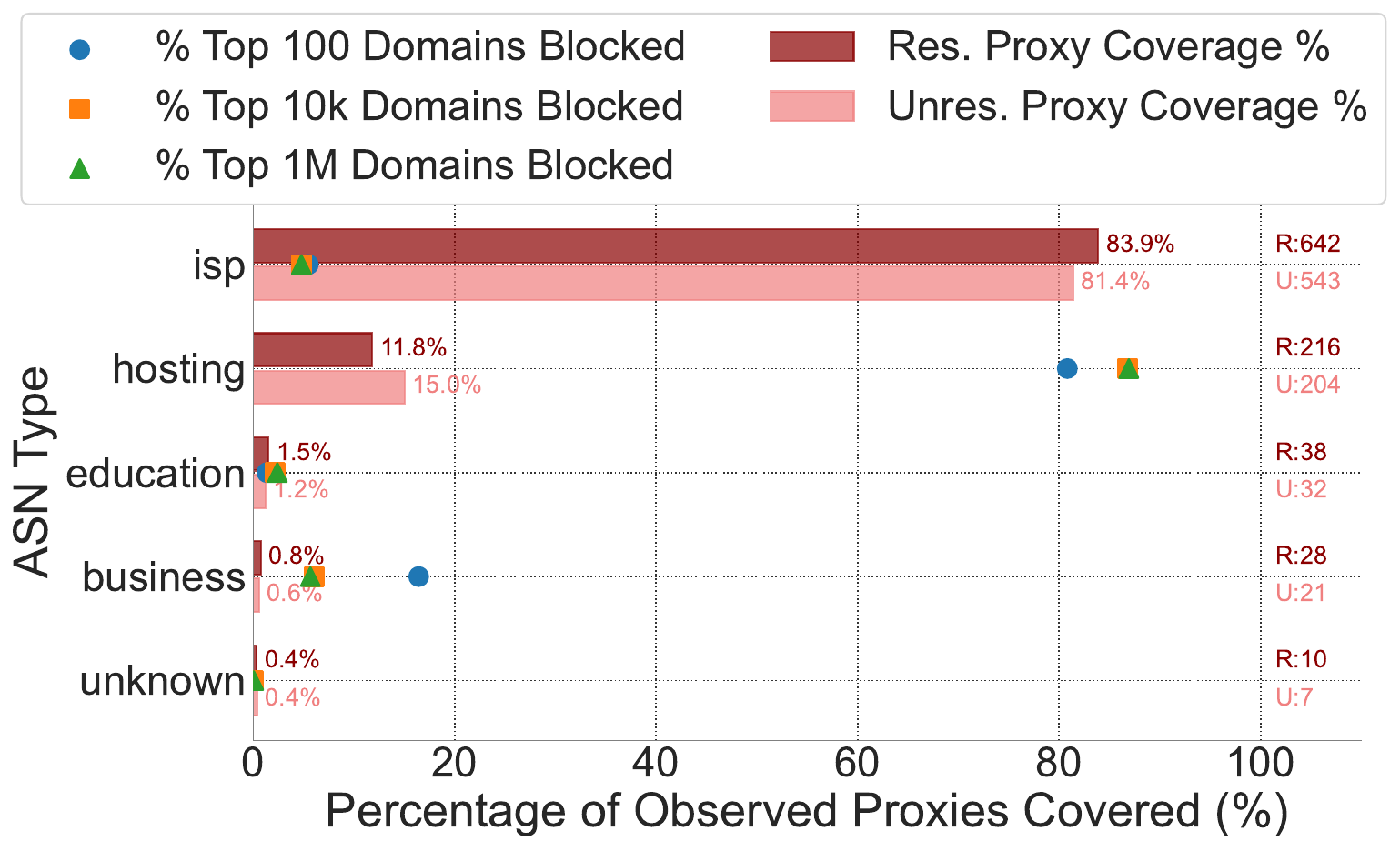}
  \caption{Blocking coverage and collateral damage by AS type. ISP ASes account for >80\% of proxies while inducing less than 10\% collateral across all domain rank bins.}
  \label{fig:blocking-asn-type-collateral}
\end{figure}

 Blocking effectiveness is also highly asymmetric across AS types: ISP networks account for more than 80\% of observed proxies while inducing very little domain collateral (<10\% across all domain ranks) as shown in Figure~\ref{fig:blocking-asn-type-collateral}, whereas hosting networks contribute far fewer proxies (<15\%) but very high collateral (>80\% across all domain ranks). The remaining classes host a comparatively minor fraction of observed proxies. Our results show that an attacker can disrupt most of the observed proxy population by focusing on ISP-heavy infrastructure, while avoiding the much steeper collateral costs associated with hosting-provider blocking.

 We note that there are other costs to blocking residential ISPs, such as blocking peer-to-peer services, that may increase collateral cost for a censor. We leave this analysis to future work.

\takeaway{More than 50\% of observed proxies can be blocked with negligible Top~100 collateral as they are in ISP networks, but the collateral costs increase sharply when the blocklist includes major cloud ASes such as Amazon.}

\begin{figure*}[t]
    \centering

    \begin{subfigure}[t]{0.33\textwidth}
        \centering
        \includegraphics[width=\linewidth]{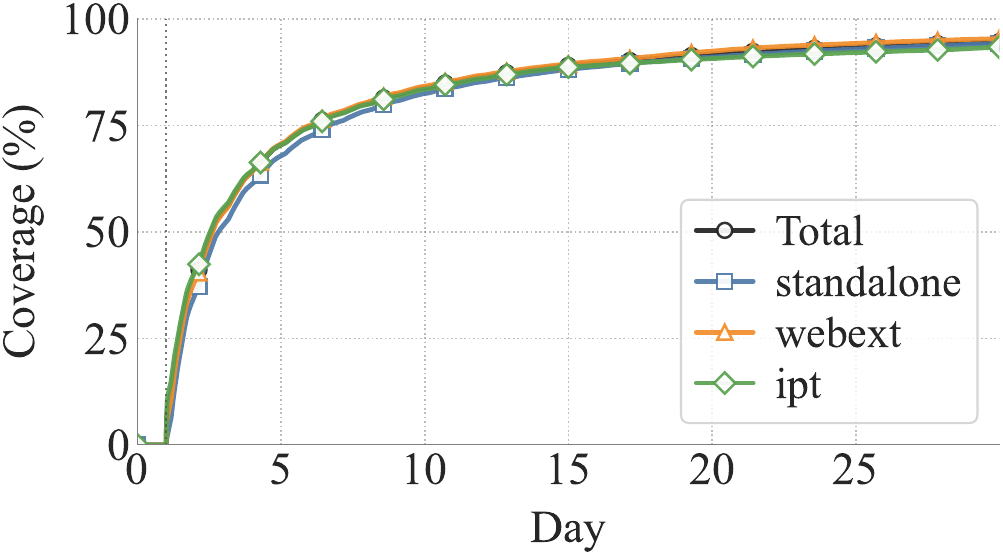}
        \caption{Default}
        \label{fig:sim-enum-default}
    \end{subfigure}\hfill
    \begin{subfigure}[t]{0.32\textwidth}
        \centering
        \includegraphics[width=\linewidth]{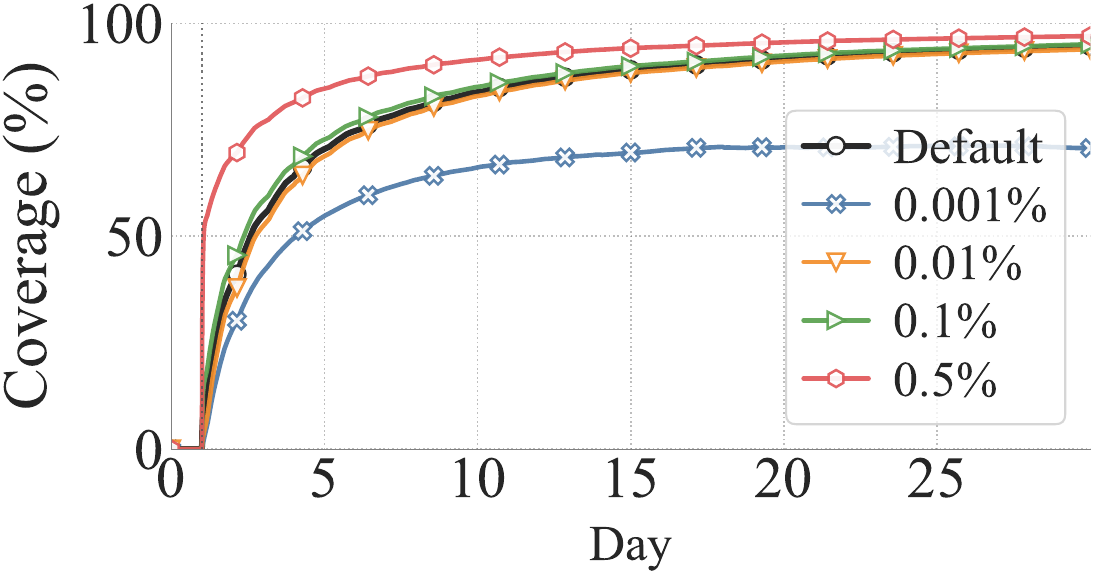}
        \caption{Attacker Fraction}
        \label{fig:sim-enum-attacker}
    \end{subfigure}\hfill
     \begin{subfigure}[t]{0.32\textwidth}
        \centering
        \includegraphics[width=\linewidth]{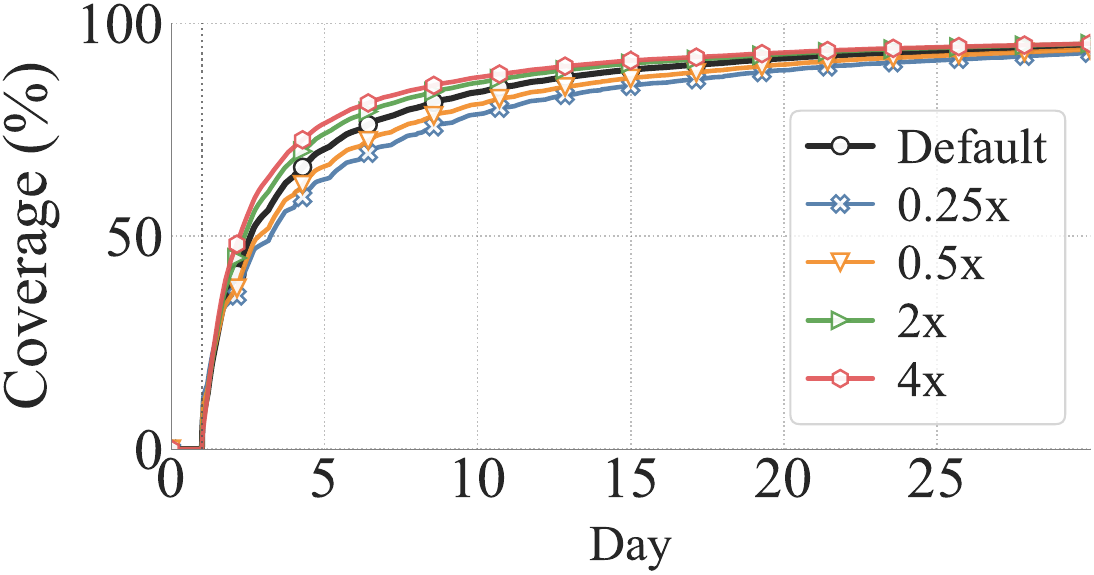}
        \caption{Churn Rate}
        \label{fig:sim-enum-churn}
    \end{subfigure}

    \caption{Cumulative enumeration coverage (percentage of total proxies polled till time $t$ that have been seen by attacker) across simulation factors. We showcase enumeration coverage per-Snowflake proxy type for the simulation settings in Appendix~\ref{sec:app-simulation-enumeration-results}.} 
    \label{fig:simulation-enumeration-all}
\end{figure*}

\subsection{Simulation Results}
\label{sec:simulation-results} 
We next discuss the results from our simulation experiments described in \S\ref{sec:simulation-experiment}, where we conduct our enumeration and blocking attacks under various settings (described in Table~\ref{tab:different-settings}). 

\subsubsection{Enumeration Attack.} 
We first describe the results of our enumeration experiments, conducted over a simulated period of 30 days. We present an overview of our simulation results in Figure~\ref{fig:simulation-enumeration-all}, where we show the cumulative percentage of all polled proxies that are enumerated by the attacker over time under the Default, Attacker Fraction, and Churn Rate parameters.

In the \textit{Default} setting, the attackers (consisting of 0.05\% of all clients) reach a mean total enumeration coverage of $94.65\% \pm 0.01\%$ by the end of our 30-day simulation, with the coverage rising from $39.22\%$ of all proxies at 48 hours to $54.96\%$ at 72 hours before converging slowly (Figure~\ref{fig:sim-enum-default}). The enumeration coverage across all proxy types are similar. 
Compared to our real-world experiments, the simulations assume a stronger attacker, with 32 malicious clients polling once per simulated second each; therefore, they yield substantially higher coverage. Because attacker–broker requests complete at the configured polling rate without transport delays or losses, the simulator does not capture reductions in effective polling throughput caused by failed or delayed requests. Enumeration coverage may therefore be higher than what an attacker with the same nominal polling rate would achieve in deployment. Therefore, we primarily use the simulation to compare how attack success changes across controlled settings.

\myparagraph{Effect of Attacker Fraction.} Enumeration coverage increases monotonically with attacker scale. Figure~\ref{fig:sim-enum-attacker} shows that with only 2 attackers (the 0.001\% setting), coverage rises much more slowly and reaches only $70.71\%$ coverage by the end of the 30-day simulation, a pattern closer to what we observe in our real-world measurements. However, the first few additional attackers produce a sharp improvement: increasing the attacker scale to 0.01\% raises final coverage to $93.88\%$, while the 0.1\% attacker setting  raises it further to $95.07\%$. Coverage continues to increase at larger attacker fractions, but most of the marginal benefit appears early in the enumeration. 

\myparagraph{Effect of Churn Rate.} Contrary to the intuition that higher proxy churn should suppress enumeration, Figure~\ref{fig:sim-enum-churn} shows slightly higher cumulative coverage as churn increases. Our investigation shows this is because higher churn increases turnover without substantially enlarging the active search space, and 32 attackers are sufficient to learn nearly all newly created proxies as they appear. In experimental runs with only 2 attackers, higher churn does reduce enumeration success. In addition, churn breaks more client-proxy assignments, returning replacement capacity to the broker pool where it can be observed again by the attackers. Thus, at this attacker scale, higher churn does not prevent enumeration, although we show later that it still reduces blocking impact.

\myparagraph{Effect of Connection Duration.} Enumeration coverage is largely insensitive to the tested changes in connection duration, as we show in Appendix~\ref{sec:30m-default}. As we describe next, however, it has a much larger impact on the blocking attack.

\takeaway{Enumeration success is primarily driven by attacker scale. Many attackers polling aggressively can quickly reach high coverage even when proxy churn is high.}

\begin{figure*}[t]
    \centering

    \begin{subfigure}[t]{0.33\textwidth}
        \centering
        \includegraphics[width=\linewidth]{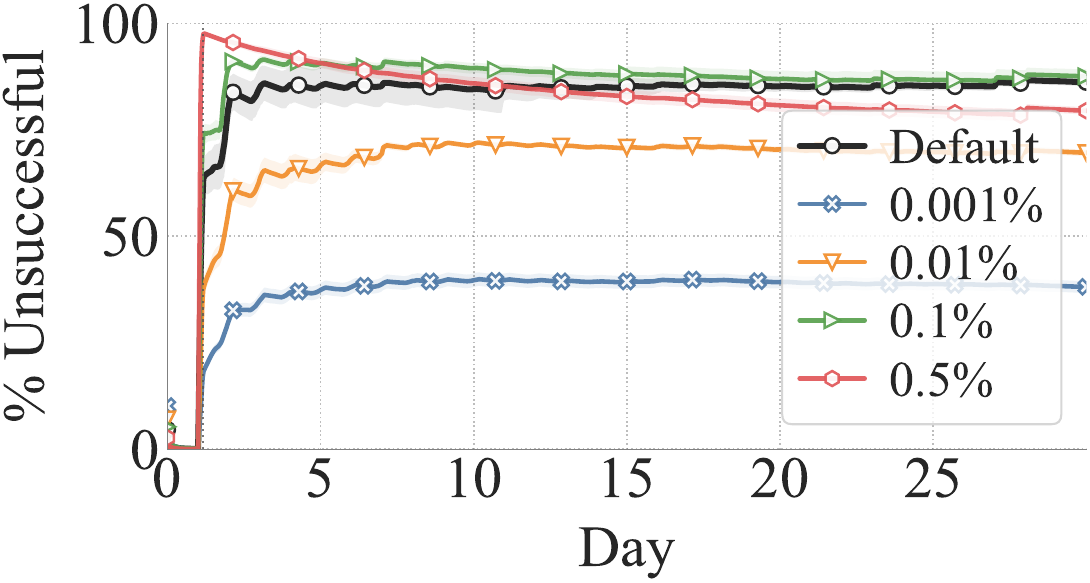}
        \vspace{0.25em}

        \includegraphics[width=\linewidth]{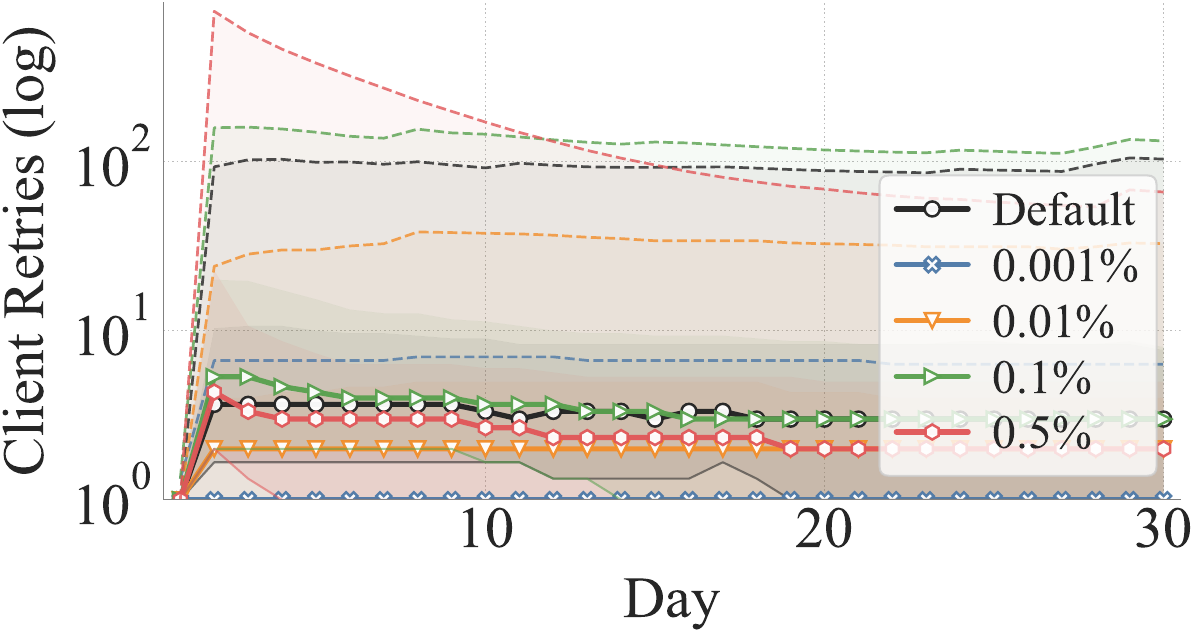}
        \caption{Attacker Fraction}
        \label{fig:sim-block-attacker}
    \end{subfigure}
    \begin{subfigure}[t]{0.33\textwidth}
        \centering
        \includegraphics[width=\linewidth]{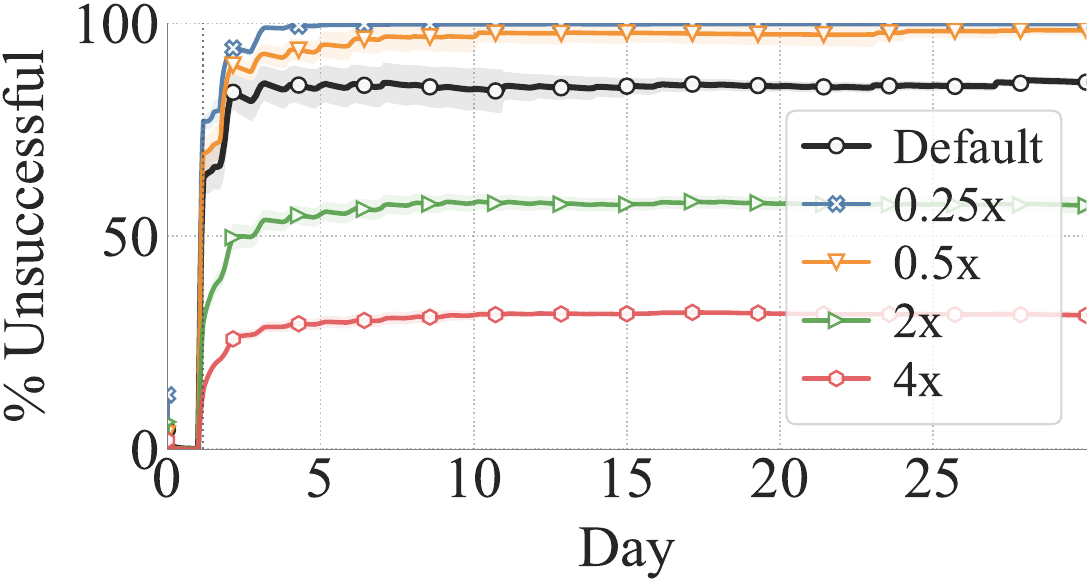}
        \vspace{0.25em}

        \includegraphics[width=\linewidth]{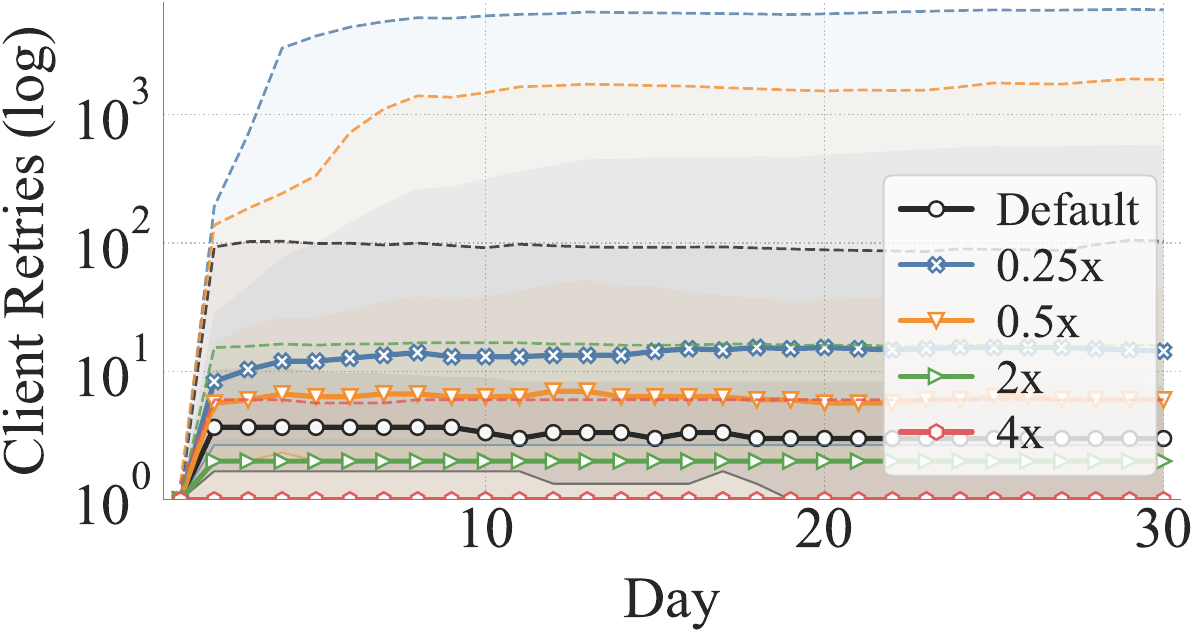}
        \caption{Churn Rate}
        \label{fig:sim-block-churn}
    \end{subfigure}
    \begin{subfigure}[t]{0.33\textwidth}
        \centering
        \includegraphics[width=\linewidth]{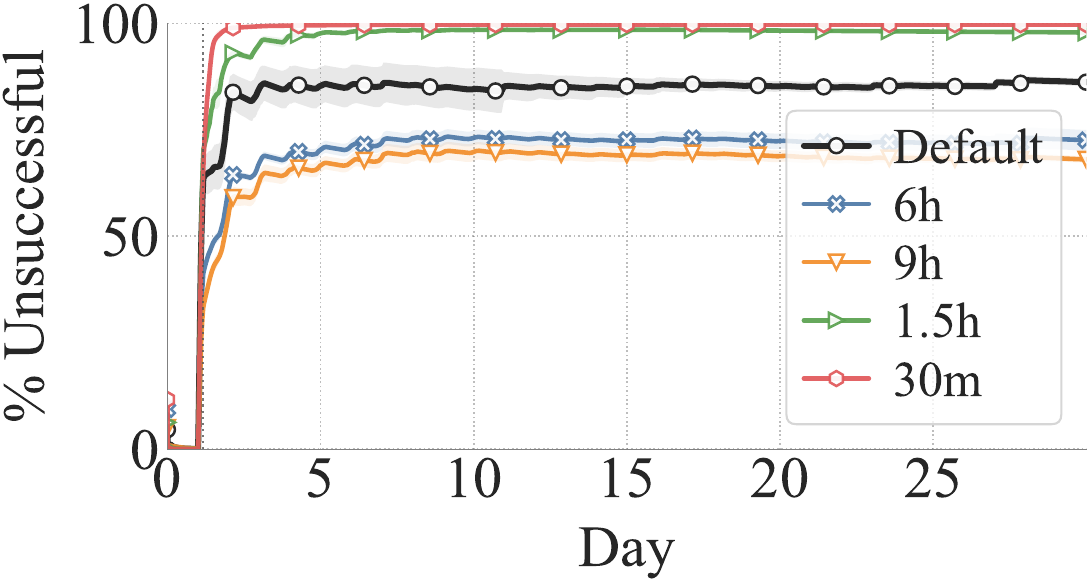}
        \vspace{0.25em}

        \includegraphics[width=\linewidth]{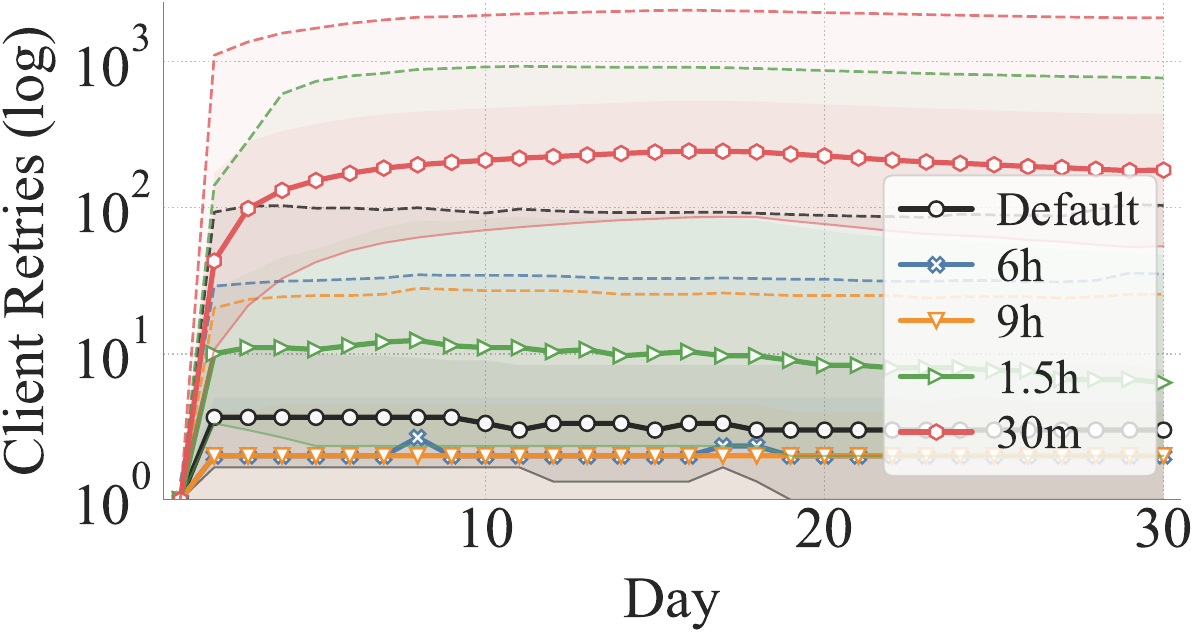}
        \caption{Connection Duration}
        \label{fig:sim-block-connection}
    \end{subfigure}

    \caption{Blocking results across simulation factors. The top plots show the cumulative percentage of client rendezvous attempts up to hour $t$ that are unsuccessful and led to a retry. The bottom plots show the distribution of average client retries among requests that required at least one retry: the lower solid line shows the 25th percentile, the marked line shows the median, the shaded bands show the interquartile and upper-tail ranges, and the dashed line marks the 99th percentile.}
    \label{fig:simulation-blocking-all}
\end{figure*}

\subsubsection{Blocking Attack.} 
Next, we explore the blocking attack, where every enumerated proxy ID is added to a blocklist and attempted matches to blocked proxy IDs fail. Figure~\ref{fig:simulation-blocking-all} summarizes client harm using two cumulative metrics: (1) the top figures show the percentage of client requests up to day $t$ that do not immediately obtain a usable proxy, either due to blocking or the proxy pool being empty; we include the latter metric since we observe that aggressive attacker polling leads proxies being popped out, which results in the same ultimate unavailability for a client, (2) the bottom figures show the distribution of client retries among requests that required at least one retry. Appendix~\ref{app:app-simulation-blocking-results} reports these blocking outcomes separately for clients with restricted and unrestricted NAT types.

In the default case (black line), most clients see the impact of the blocking, but the median client remains able to recover a working proxy at a limited cost. By day 30, the cumulative unsuccessful request rate reaches $86.16\% \pm 0.74\%$, while among requests that required retry, the median request needs 3 retries, the 75th percentile remains below 8, and the 99th percentile exceeds 100 retries. Thus, most affected requests still recover after a small number of retries, but the tail becomes expensive: with a three-second retry interval, 200 retries correspond to $\sim$10 minutes of user wait time.

\myparagraph{Effect of Attacker Fraction.} Increasing attacker fraction generally increases blocking harm. The percentage of unsuccessful requests rises from $38.12\%$ at 0.001\% attackers to $86.16\%$ in the default setting and $87.53\%$ at 0.1\%, while the median number of retries increases from 1 to 3. The 0.5\% case is non-linear over time: our inspection of the logs shows that, at this scale, failures shift away from long blocked-proxy retry chains and toward faster \texttt{No Proxies} outcomes.

\myparagraph{Effect of Churn Rate.} Churn has the strongest effect on blocking. Reducing churn to $0.25\times$ the default increases the cumulative unsuccessful request rate to $99.80\%$, with a median retry count of 14, a 75th percentile of 571, and a 99th percentile above 6,500 retries; even at $0.5\times$ churn, the unsuccessful request rate remains $98.30\%$. By contrast, increasing churn reduces blocking harm: at $2\times$ churn, the unsuccessful request rate falls to $57.20\%$ with a median of 2 retries, and at $4\times$ churn it falls further to $31.42\%$ with a median of 1 retry. Thus, even though higher churn does not reduce cumulative enumeration coverage, it substantially weakens the blocking attack by reducing how long blocked proxies remain useful to the attacker.

\myparagraph{Effect of Connection Duration.} Shorter client-proxy sessions substantially worsen blocking. At 30m~$\pm$~5m, the cumulative unsuccessful request rate reaches $99.70\%$, while the median retry count increases to 200 and the 99th percentile exceeds 2,100 retries. Even at 1.5h~$\pm$~15m, the unsuccessful request rate remains $97.82\%$. By contrast, longer sessions reduce blocking pressure: at 6h and 9h, the unsuccessful request rate falls to $73.21\%$ and $68.57\%$, respectively, and the median retry count drops to 2. We see that shorter sessions increase competition for working proxies from returning clients.

\takeaway{More attackers, lower churn and shorter sessions make blocking extremely costly for clients, while higher churn reduces blocking. Overall, proxy churn remains Snowflake's primary defense against IP-based blocking attacks.}

\begin{figure}[t]
    \centering
    \includegraphics[width=0.49\linewidth]{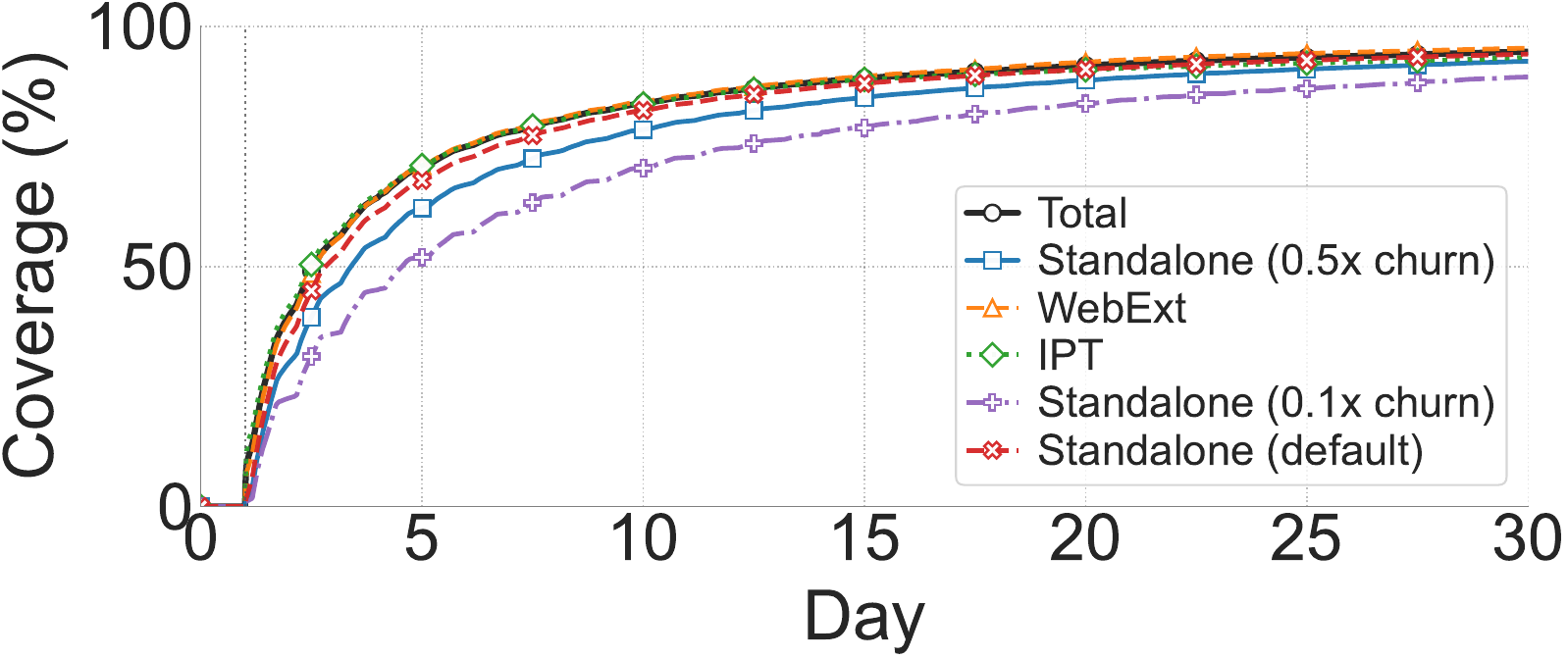}
    \includegraphics[width=0.50\linewidth]{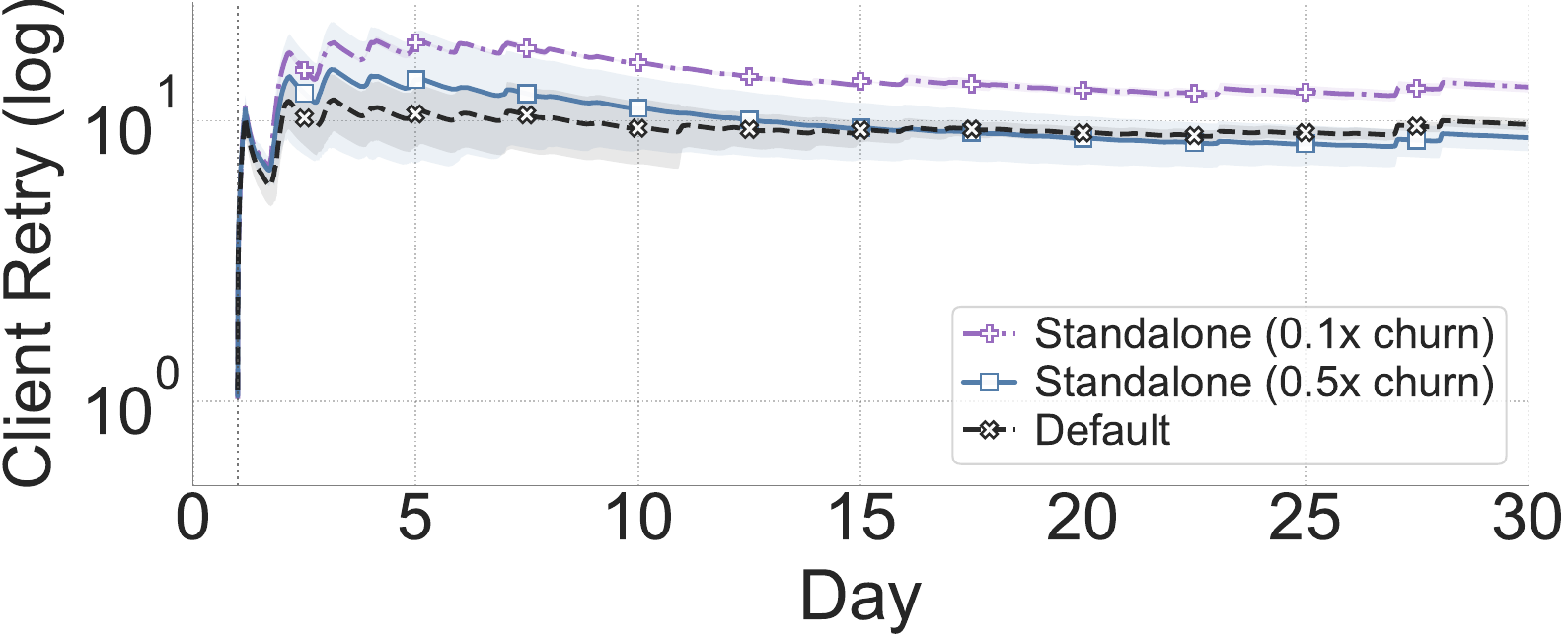}
    \caption{Although standalone proxies constitute only a
small fraction (4\%) of the proxy pool, when standalone proxies churn at the 1/2 and 1/10 of the default rate, the enumeration of standalone proxies is apparently less than default case and reducing their churn
rate significantly increases the number of client retries.}
    \label{fig:standalone-half-churn}
\end{figure}

\subsubsection{Case Study: Slower Churn Standalone Proxies.} 
\label{sec:standalone-case-study}
Standalone proxies typically reside for long time periods on larger servers which do not often change their IP addresses and other network settings. Accordingly, we conduct experiments with standalone proxies churning at one-half and one-tenth of the default churn rate. The results are shown in Figure~\ref{fig:standalone-half-churn}. As the churn of the standalone proxies decreases, the attacker can only enumerate a smaller fraction of total proxies, which aligns with our earlier comparison of churn rate in Figure~\ref{fig:simulation-enumeration-all}. At one-tenth of the default churn rate, enumeration coverage reaches 51.95\% by day 5 and 89.43\% by day 30, compared with 67.88\% and 94.21\%, respectively, under the default churn rate. Our investigation shows that a higher churn rate will introduce more new proxies into the broker heap, and the new proxies that serve no clients will be ranked first, which the attacker is more likely to find. However, from the perspective of clients, the average retries for a client in the case of a blocking attacks would be higher if standalone proxies churn more slowly, which aligns with earlier blocking results in Section~\ref{sec:simulation-results}. Relative to the default case, reducing churn to one-half increases the mean number of client retries by 0.5, while reducing it to one-tenth increases the mean by 5 retries.

\takeaway{Slower churn reduces attackers' ability to enumerate standalone proxies, but increases the number of retries clients experience under blocking attacks.}

\section{Practical Malicious Proxy Attack}
\label{sec:malicious-proxy}
In addition to our malicious client attacks on Snowflake, we identify another practical attack through our investigation, where an adversary acts instead as a \textit{malicious Snowflake proxy}. Since Snowflake does not impose any rate limits on Snowflake proxy polls, a malicious actor can flood the broker's proxy heap structure with malicious proxies that will be then served to many clients. As noted in~\S\ref{sec:snowflake-background}, the broker currently sorts proxies in its priority heap according to a self-reported number of clients currently served by a proxy. A malicious proxy can consistently poll the broker with the number of clients set to $-1$, which would result in it being at the front of the priority heap because the smallest number of clients a benign proxy would claim is 0. 

We conducted a case study of such a malicious actor using our simulated environment, polling every second with two malicious proxies (one with a restricted NAT and the other unrestricted) that had the number of clients set to $-1$, and recording how many client connections were made to the malicious proxies over time. We observe that our malicious proxies are consistently offered to an average of 19.8\% of all client requests across a 15-day simulation period. A real-world attacker can scale this attack even further by either polling faster or by polling with multiple malicious proxies. 

We informed the Snowflake maintainers promptly regarding this vulnerability, who were already aware of the problem~\cite{tor-malicious-proxy}. Although the malicious proxy attack is a known attack surface, we evaluate its effect quantitatively and show that the attack is particularly harmful for Snowflake as it allows self-reporting client loads. A powerful malicious proxy can use this vulnerability to  disproportionately learn Snowflake client IP addresses. These identifiers could subsequently be correlated with other network observations to support targeted monitoring or blocking. A malicious proxy could also refuse to relay traffic, forcing affected clients to retry and creating an availability attack. These results motivate reducing the influence that proxies have over their position in the broker’s selection process. Our proposed defenses to enumeration and blocking attacks discussed next in \S\ref{sec:mitigations} will help reduce the impact of malicious proxy attacks.

\takeaway{Two aggressively polling malicious proxies receive almost 20\% of client assignments in our simulation, giving them a disproportionate view of connecting clients.}

\section{Defenses and Mitigations}
\label{sec:mitigations}
While the churn of the deployed Snowflake network offers some resistance to practical enumeration and blocking attacks, we discuss some modifications to Snowflake based on our simulation results that we believe will further increase Snowflake's censorship resistance.  In our simulations, a sufficiently active attacker can enumerate most of the proxy population even under substantial churn, but the downstream harm of blocking depends on how often the broker exposes the same proxies and how long enumerated proxies remain useful to the attacker. We therefore organize defenses around three goals: limiting the attacker's share of rendezvous requests, reducing broker-amplified exposure of stable proxies, and making blocklists decay faster.

\subsection{Rate Limiting Client Polls}
\label{sec:client-rate-limiting}
A natural defense against enumeration is to rate-limit rendezvous attempts from client IP addresses, since attack success requires repeatedly querying the broker for proxy assignments. Our attacker-fraction simulations support this intuition (Figure~\ref{fig:sim-enum-attacker}): with only 0.001\% attackers, final enumeration coverage is substantially lower than in the default setting. %

However, based on our experience working with Snowflake, it is difficult to fairly and accurately limit client rendezvous attempts based on IP addresses. Honest clients often need several rendezvous attempts before obtaining a usable proxy. NAT incompatibilities, proxy churn, packet loss, and transient rendezvous failures can all require a client to request a new proxy. Multiple users may also share the same public IP address, and many broker rendezvous channels involve intermediaries that forward client requests. Although client IPs may sometimes be recovered from headers or WebRTC offers, these signals are not always reliable and may be spoofed by an attacker. That is why Snowflake has explored but ultimately decided not to limit client polling~\cite{client-rate-limit-issue}. 

A more robust future direction is to rate-limit around a privacy-preserving notion of client effort rather than raw IP address. For example, future work can explore requiring lightweight proof of work, anonymous credentials, or broker-issued retry tokens that allow honest clients to recover from failed matches while making sustained high-rate enumeration more expensive.  Even with these techniques, we maintain that the unpredictability of the live Snowflake network makes balancing the enumeration of proxies with availability to honest clients difficult work. 

\subsection{Broker-Scheduled Proxy Polling}
\label{sec:proxy-fairness}
\begin{figure}[t]
    \centering
    \includegraphics[width=0.49\linewidth]{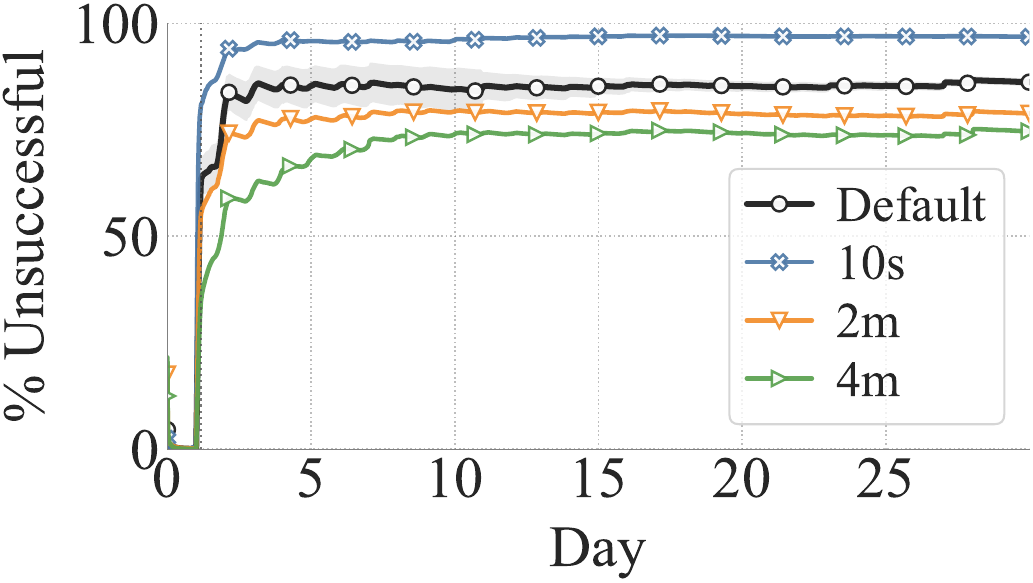}
    \includegraphics[width=0.50\linewidth]{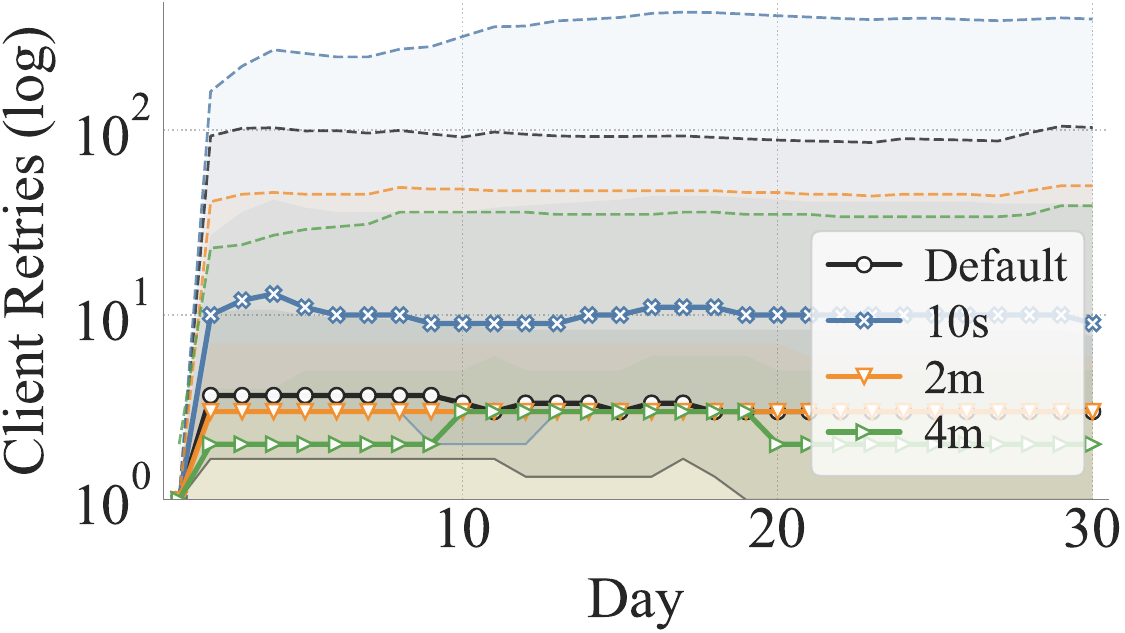}
    \caption{Blocking results for the equal polling defense. Plots are generated similarly to Figure~\ref{fig:simulation-blocking-all}. The left figure shows the cumulative percentage of client requests that are unsuccessful and the right figure shows the distribution of client retries when at least one retry is required.}
    \label{fig:equal-polling-blocking}
\end{figure}

Our results show that the impact of enumeration and blocking attacks is greater because the proxies most likely to be enumerated are also stable proxies disproportionately more likely to be distributed to users. The malicious-proxy experiment illustrates the same mechanism from the opposite direction: just two malicious proxies can receive a disproportionate share of client assignments, appearing in roughly 20\% of client matches on average.

To tackle this issue, we propose broker-scheduled proxy polling at equal rates. Instead of allowing high-capacity or aggressive proxies to poll as frequently as they choose, the broker can, ideally dynamically, assign each proxy a next poll time and reject or delay polls that arrive too early.  If a \texttt{standalone} proxy and a \texttt{webext} proxy are both under capacity and online, this defense makes it equally likely that they be matched with a client within a fixed time frame. We recommend the equal proxy poll rate be kept as low as necessary to satisfy client demand. This ensures that when low capacity proxies are available, they are not crowded out by high capacity proxies that are polling more frequently than necessary. 

We perform a simulation round to showcase the effect of equal proxy poll rates. We consider all proxies polling at equal rates of 10 seconds, 120 seconds, and 240 seconds, and measure the impact of our default blocking attacker. As shown in Figure~\ref{fig:equal-polling-blocking}, longer polling rates reduce both the fraction of unsuccessful benign requests and the upper tail of client retries. We posit that even lower polling rates can further reduce blocking. 

In deployment, we recommend a \textit{dynamic} broker-scheduled polling interval that adjusts according to client demand. Beyond reducing blocking success, this design also: (1) reduces the impact of malicious proxies, since they cannot crowd out honest proxies, and (2) creates a path towards removing self-reporting of client load by allowing the broker to maintain its own records of proxy availability and assignments. 
 The Snowflake team acknowledged and deployed a fix for the malicious proxy vulnerability by enforcing the self-reported number of clients to be at least 0 in April 2026 [\textit{Reference Blinded}]. This closes the specific negative-client-count exploit evaluated in our attack, although the broker must still rely on proxies to report their loads because it cannot observe when peer-to-peer connections fail or terminate. Informed by our results, Snowflake has also progressively implemented broker-scheduled proxy polling and broker-side rate limits [\textit{Reference Blinded}]. In March 2026, they introduced a ``NextPoll'' field through which the broker communicates the next permitted polling time to a proxy. In August 2026, they added broker-side enforcement that rejects polls arriving before the permitted time. At the time of writing, the enforcement code has been merged into the main Snowflake codebase, while work continues on selecting and dynamically adjusting the polling intervals used in deployment.

\subsection{Increase Proxy Churn}
We found in \S\ref{sec:simulation-results} that increasing the churn rate of proxies significantly reduced  the number of attempts needed for a client to be matched with an unblocked proxy. While the churn of proxies that self-report to the Snowflake broker as \texttt{standalone} is lower than for self-reported \texttt{webext} proxies, the usage of Spot VMs similar to SpotProxy~\cite{kon2024spotproxy} could increase proxy churn without sacrificing the benefits of unrestricted NATs that come with VPS hosting. %
Running more high churn proxies on cloud providers may also help recover the network in the case of AS-level blocks of consumer ISPs described in \S\ref{sec:real-world-blocking}.

\section{Discussion}
\label{sec:discussion}
\myparagraph{Implications for Proxy-Based Circumvention Systems.}
Our primary contribution is an empirical evaluation of how a deployed circumvention system’s design translates into practical enumeration and blocking risk. The results show that the nominal size of a proxy pool is not, by itself, an adequate measure of enumeration resistance. Resilience also depends on which proxies the broker repeatedly exposes, how much client capacity those proxies provide, how quickly observed addresses become stale, and how concentrated the proxies are across networks. In Snowflake, stable high-capacity proxies are exposed more often, making partial enumeration useful. This lesson may extend to other proxy-based tools such as Psiphon~\cite{psiphon-webpage}, though the risk depends on each system's discovery mechanisms, matching policies, and diversity.

\myparagraph{Ethical Measurement of Live Circumvention Systems.}
A central lesson of this work is that ethical constraints must carefully shape the measurement design, especially when performed with a deployed system. Our safety-first design reduced the strength and duration of our live attack, but they were necessary to avoid harming a deployed system used by at-risk users. We believe this combination of bounded real-world measurements, close coordination with system operators, and simulation for evaluating more aggressive settings provides a useful model for future studies of deployed circumvention infrastructure.

\myparagraph{Limitations and Future Work.}
Our study has several limitations. First, we do not perform real-world blocking; instead, we estimate blocking impact from subsequent broker observations and controlled simulations. Second, our collateral analysis focuses on web-domain overlap and does not fully capture the broader social, economic, or connectivity costs of blocking residential or access networks. Third, as discussed in \S\ref{sec:simulation-experiment}, our simulator preserves the broker interactions relevant to enumeration and blocking, but abstracts away full WebRTC/Tor traffic, packet loss, and latency. Besides, we also simplify the churn model to maintain tractability and avoid overfitting to certain real-world scenarios. Although our assumptions are informed by public Snowflake metrics, the simulation may not capture all real-world noise and network behavior, so our simulation results should be interpreted primarily as comparative trends rather than precise absolute estimates.
Finally, our results reflect Snowflake’s current design and deployment, and may change as defenses are integrated, which can be evaluated by future work using our open-source tools (refer Appendix~\ref{sec:open-science}).

\section{Related Work}
\label{sec:related-work}
Previous research on circumvention systems has primarily focused on the detection, fingerprinting, and blocking of circumvention tools based on features of their \textit{observable network traffic}. Prior work has shown that censors can distinguish Tor traffic, pluggable transports, and other circumvention tools using packet sizes, timing patterns, and handshake fingerprints, enabling targeted blocking~\cite{fifield2015blocking,fifield2016fingerprintability,winters12foci,winters12foci,wails2024precisely,geddes2013cover,houmansadr2013parrot,frolov2019use}. Similar measurement studies have demonstrated large-scale enumeration and blocking of VPN endpoints, highlighting the feasibility of adversaries discovering VPN infrastructure~\cite{li2022vpn,xue2022openvpn}. Censors have also demonstrated the ability to identify and disrupt fully encrypted circumvention traffic by exploiting flow-level characteristics and protocol behaviors, as seen in China's blocking of end-to-end encrypted protocols~\cite{gfw-fully-encrypted}.

Another line of research tries to identify \textit{whether proxies can be actively probed} and how to protect against such probing. Work on active probing detection in Tor bridges shows that censors routinely scan for and block proxy endpoints using proxy response features~\cite{EnsafiHiddenChina,dunna2018analyzing}. This work motivated a new class of circumvention protocols resistant to active probing attacks, such as obfs4 and ScrambleSuit~\cite{winter2013scramblesuit,obfs4}. Smits~\etal proposed BridgeSPA, which uses single-packet authorization to authenticate clients before revealing Tor bridge information, reducing the risk of bridge discovery by censors~\cite{smits2011bridgespa}.  When Durumeric~\etal introduced Zmap, they performed Internet-wide scans on ports 443 and 9001 to locate 86\% of hidden Tor bridges~\cite{zmap}. Frolov~\etal used identifying features of TCP behavior to detect probe-resistant proxies~\cite{frolov2020detecting}. Compared to Internet-wide Tor Bridge discovery~\cite{zmap}, our work does not employ Internet-wide probing for proxies: rather, \textit{we actively enumerate proxies through the centralized Snowflake broker}.  

Closest to our work, Ling~\etal performed a large-scale empirical evaluation of Tor bridge discovery in 2013 by probing HTTPS and email servers and deploying malicious Tor middle relays~\cite{ling2013tor}.
Compared to this Tor bridge enumeration work, the Snowflake ecosystem presents several new challenges: Snowflake proxies are ephemeral and the broker relies on self-reported values to prioritize distribution, while Tor bridges have long-term identifiers in the form of fingerprints that are fetched and verified from special directory authorities. In addition, the scale of Tor bridges ($\sim$2.5K) is much smaller than Snowflake ($>$100K). Snowflake addresses the proxy enumeration problem in a fundamentally different way: by reducing the barrier to running a proxy and therefore offering a larger and more churning pool of IP addresses. Besides, some countermeasures proposed for Tor bridge discovery like human interaction (CAPTCHA)~\cite{ling2013tor} are also not suitable for Snowflake. These differences  warrant separate study for Snowflake proxy enumeration and blocking.

Research on detecting and blocking Snowflake has focused on the distinguishability of Snowflake requests from regular Internet traffic. Xie~\etal~\cite{10152736} used machine learning on packet size and timing features to distinguish Snowflake rendezvous requests from five other types of web requests.  MacMillan~\etal showed how the DTLS features of Snowflake WebRTC traffic can be distinguishable from other services that use WebRTC~\cite{macmillan2020evaluating}.
The Snowflake development team also reported several methods used by countries to block Snowflake, though sometimes unintentionally, and subsequently patched these vulnerabilities~\cite{bocovich2024snowflake}. For example, they discovered that the Russian blocking of Snowflake in 2021 was a result of DTLS fingerprinting and patched the issue. Our work differs fundamentally from this previous research: \textit{rather than considering features for traffic fingerprinting, we consider a practical attacker who attempts to enumerate and block proxies in the system through active probing of a centralized broker.} 

Also relevant to our work is the line of work on proxy distribution strategies~\cite{nasr2019enemy,kon2024spotproxy,fares2026game,douglas2016salmon,wang2013rbridge,mccoy2011proximax}. Nasr~\etal~\cite{nasr2019enemy} modeled proxy distribution as a game between circumvention system operators and the censors,
and use game theory to derive the optimal strategies of each of
the parties. Their work showed through simulations that there are optimal strategies that can help systems like Tor distribute proxies in a censorship-resistant manner.  Kon~\etal~\cite{kon2024spotproxy} proposed SpotProxy, a practical circumvention system that uses frequently changing cloud-based proxies on spot instances to reduce operational cost while enabling rapid migration across proxies and IP addresses. They evaluate SpotProxy’s circumvention performance using Nasr~\etal’s game-theoritic simulation framework~\cite{kon2024spotproxy}.  More recently, Fares~\etal~\cite{fares2026game} revisited Nasr~\etal’s framework in light of modern proxy-based circumvention systems such as Snowflake. They extend the simulation model to capture ephemeral browser- and mobile-based proxies, NAT restrictions, and traffic-analysis-based enumeration strategies, and show that earlier ``optimal'' strategies do not necessarily remain optimal in these newer settings. Our work differs from this line of research in three important ways: \textit{we conduct real-world measurements against a live circumvention system, we build large-scale simulations grounded in observed real-world behavior rather than purely abstract models, and we evaluate the collateral damage induced by network blocking.}

\section{Conclusion}
We presented the first systematic empirical evaluation of practical enumeration and blocking risk in the deployed Snowflake ecosystem. By combining a bounded 48-day live measurement with full-scale controlled simulation, we evaluated two assumptions underlying Snowflake’s blocking resistance: that adversaries cannot comprehensively enumerate its dynamic proxy pool and that blocking observed proxies would impose substantial collateral damage. Our results show that malicious clients can quickly enumerate stable, high-capacity proxies and that AS-level blocking can be effective with limited collateral cost for adversaries. We also highlight that attacker scale and proxy churn play a significant role in attack success. We discuss and evaluate practical mitigation that can reduce attacker visibility and blocking impact.

\begin{acks}
We thank the anonymous reviewers for their constructive feedback. We also thank Felicity Shih, meskio, Shelikhoo, Lindsey Tulloch, David Fifield and other members of the Snowflake and RANDLab research groups for their insights. This work was supported by a UC Santa Cruz COR Large Grant. 
\end{acks}

\bibliographystyle{ACM-Reference-Format}
\bibliography{main}
\section*{Generative AI Usage}
\label{sec:genai-usage}
We acknowledge the use of Generative AI tools such as ChatGPT and Copilot, primarily for assisting with the development of our Snowflake simulation tool. The core design and initial implementation of the tool, including modifications to the original Snowflake broker code, were carried out manually by the authors. Generative AI coding tools were used primarily to assist with performance optimizations (e.g., abstracted client/proxy implementations and more efficient logging strategies), as well as for documentation and generating plots. All AI-assisted contributions were guided by the authors and subsequently thoroughly manually validated to ensure correctness and reliability.
We also used generative AI tools for grammar and writing improvements. In all cases, all generated content was carefully reviewed by the authors.

\appendix
\section{Open Science}
\label{sec:open-science}
We provide three artifacts to support evaluation of this paper's core contributions and encourage reproducibility.

\myparagraph{Real-World Prober Code \& Data.} For conducting our real-world experiments (\S\ref{sec:real-world-measurements}), we instrumented a modified Snowflake client and performed real-world enumerations, which is available here: \url{https://github.com/r-andlab/snowflake-enumeration/tree/main/real-world}. The code is based on the original Snowflake client code, and it contains our modification to act as a probing client. We also provide aggregate real-world measurement data from the 48-day enumeration study. Raw proxy IP addresses and keyed hashes are withheld to protect the privacy of volunteer proxy operators, as described Appendix~\ref{sec:ethics}.

\myparagraph{Simulation Code \& Data.} We open-source our Snowflake simulation code and data at: \url{https://github.com/r-andlab/snowflake-enumeration/blob/main/general-simulation/broker/README.md} for general simulation (\S\ref{sec:simulation-experiment}) and \url{https://github.com/r-andlab/snowflake-enumeration/blob/main/malicious-proxy-simulation/broker/README.md} for the malicious proxy scenario (\S\ref{sec:malicious-proxy}). The simulation instruments the real-world Snowflake broker~\cite{snowflake-code} with a simulation harness, and we provide detailed instructions on operating the simulations. The malicious proxy scenario adds 1 restricted malicious proxy and 1 unrestricted malicious proxy to the simulation. The repositories include a link to our simulation data.

\section{Ethical Considerations}
\label{sec:ethics}
Our experiments described in \S\ref{sec:real-world-measurements} and \S\ref{sec:simulation-experiment} were shaped by (1) a formal safety review through the Tor Research Safety Board~\cite{tor-safety-research-board}, (2) repeated consultation with Snowflake developers, and (3) guidelines outlined by the Tor Project, prior Tor research, and the Menlo Report~\cite{torproject,jansen2021once,bauer2011experimentor,jansen2011shadow,soghoian2011enforced}. Before conducting live experiments, we submitted a detailed proposal to the Tor Research Safety Board describing our objectives, anticipated benefits, potential risks, and experimental methodology. Following their helpful reviews, we engaged in several rounds of constructive discussions and feedback with the Snowflake developers that informed and refined our final experimental design and longitudinal measurements.

The primary benefit of our study was an assessment on whether censors could effectively enumerate and block proxy IP ranges without incurring significant collateral damage, and to develop actionable recommendations for improving Snowflake's resistance to blocking. We identified four central risks that were relevant to our measurements. First, repeated client-like probing could \textit{increase broker load} causing technical disruption to the Snowflake broker. Second, a prober that behaved too aggressively could \textit{consume proxy offers} that would otherwise have gone to real users. Third, collecting raw proxy endpoint IP addresses could \textit{expose sensitive information} about volunteer-operated proxies. Fourth, continuing longitudinal probing during periods of unusually high demand could impose risk precisely when the system was under the \textit{greatest operational pressure}.

We designed our experiments carefully to minimize all the above risks. First, to limit the possibility of technical disruption, we conduct most resource-intensive analyses in simulation rather than on the live Snowflake broker, and restrict our real-world measurements to only two bounded, low-rate probers. Our probers were configured to wait at least one second after a completed probe, although in practice network delays meant that the fastest sustained  probing rate we achieved was one probe every six seconds, and our average probing rate was once every 10 seconds. Relative to Snowflake's historical load of 400--700 client polls every 5 seconds~\cite{snowflakeprometheus}, we expect our measurements to impose minimal additional burden on the broker. This conservative design necessarily limits our ability to model stronger adversaries, such as nation-states operating many malicious clients at higher rates, but we believe these protections are necessary to reduce risk to the broker, proxies, and users. 

Second, we further limit our measurements to only receiving proxy offers from the broker and never completed connections to the proxies themselves \ie we never open a connection and data channel with the enumerated proxy. As a result, the proxies we enumerated remained available for legitimate clients once they re-polled, typically within less than two minutes.

Third, we never store the raw proxy IP addresses on disk. However, we still need to store a representation of the IP address in order to distinguish duplicates and newly enumerated proxies. We use a key generated randomly inside our probing code to generate a keyed hash of the IP address, and only store the hashed IP address and their corresponding ASNs (calculated using RouteViews PrefixtoAS mappings~\cite{routeviews-prefix2as}) on disk. The generated key is only kept in RAM and discarded when the program exits. While this caused some technical difficulty---for instance, not being able to resume experiments once stopped due to the loss of the key---we believe this is an essential step to prevent the accidental leak of proxy IP addresses. Moreover, we only publish aggregated open-source datasets rather than the individual hashes and timestamps, in order to prevent any correlation attacks (refer to artifacts in Appendix~\ref{sec:open-science}). 

Finally, we kept in close communication with Snowflake developers and the broker operators throughout our real-world measurements to ensure that our experiments did not impose additional load during periods of elevated demand.  Our initial study design involved probing longitudinally for 6-12 months. However, about one month into the study, Internet shutdowns in Iran led to increased volatility and demand for proxies~\cite{snowflake-june-2025,snowflake-iran-june-2025,net4people-iran-june-2025}. In response, we chose to end the live measurements after capturing the system's initial behavior under load, rather than continue probing through a period when legitimate users faced heightened need. Although this decision reduced the duration of our longitudinal real-world analysis, we made it deliberately to minimize burden on the Snowflake ecosystem and to maximize the likelihood that clients in the affected region could obtain a working proxy.

During our analysis, we also identified a vulnerability where malicious proxies could receive a large share of client assignments, affecting privacy and availability for real Snowflake clients (\S\ref{sec:malicious-proxy}). We promptly disclosed this vulnerability to the Snowflake maintainers, sharing the attack mechanism and supporting simulations. Although the Snowflake team had already recognized malicious proxies as a possible concern, they had been considered lower priority because they are outside Snowflake's primary threat model~\cite{tor-malicious-proxy}. We emphasized the importance of addressing this issue given the success of the attack in our simulations, and discussed several mitigation mechanisms (\S\ref{sec:mitigations}).

We also note that publishing this work creates both benefits and risk. By describing the enumeration procedure and identifying stable proxies and network-level concentrations, the paper could reduce the effort required for a censor to enumerate or block Snowflake infrastructure. Releasing our measurement and simulation tools may further make these attacks easier to study. We cannot eliminate this risk. We nevertheless believe that the publication and release of artifacts is justified because the attacks use public Snowflake interfaces available to ordinary clients and proxies and could therefore be independently discovered and conducted by a capable adversary, including a well-resourced nation-state censor. Our responsible disclosure allows Snowflake maintainers to assess the vulnerabilities and begin implementing mitigations before wider release. Public analysis also enables the research and circumvention communities to evaluate the underlying assumptions, reproduce our aggregate findings, and develop additional defenses.  We also note that related privacy and censorship research has similarly weighed dual-use risks against the benefits of reproducibility and public scrutiny~\cite{wails2024precisely,cherubin2022online}. 
Overall, in balancing these risks and benefits, we concluded that coordinated disclosure and publication of privacy-preserving results offered greater long-term benefit to Snowflake users than leaving the weaknesses undocumented.

\section{Real-World Blocking Attack: Network Blocking Strategies}
\label{sec:app-real-world-blocking-strategies}
\begin{figure}[!t]
  \includegraphics[width=0.99\linewidth]{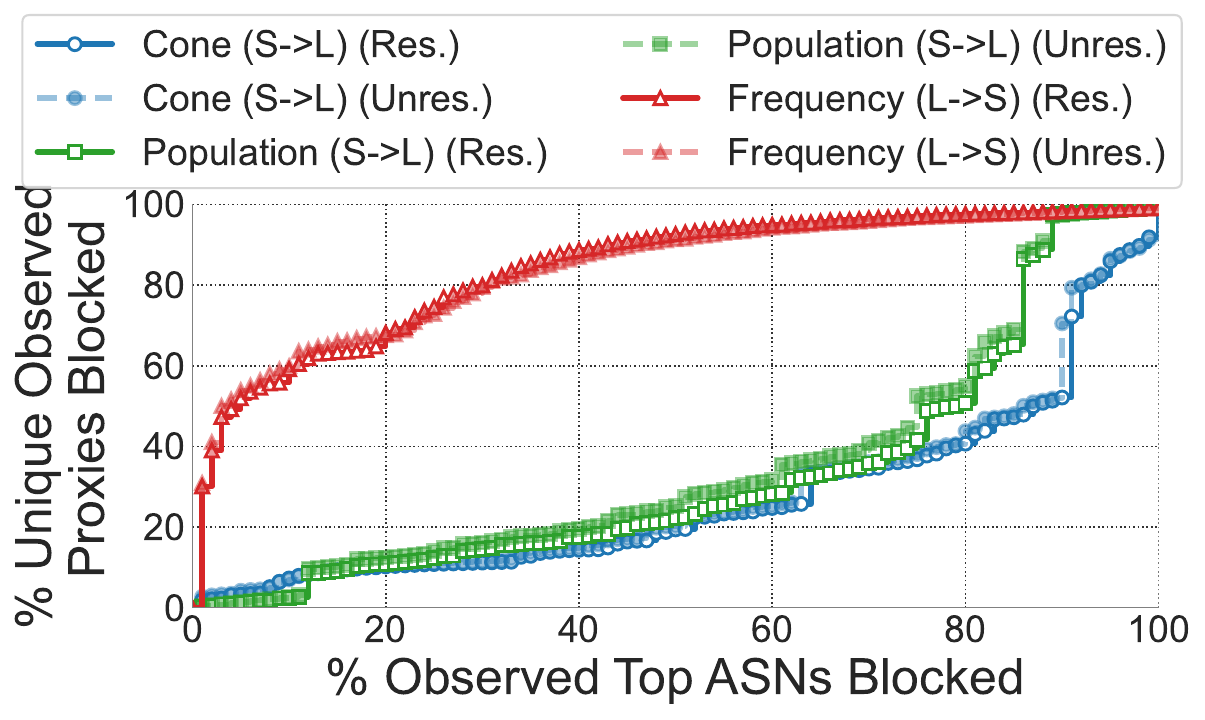}
  \caption{Observed Snowflakes blocked under different ASN blocking orderings. Blocking ASes by decreasing order of observed Snowflake frequency is substantially more effective than blocking by AS cone size or country population. The Top ASNs are ordered by smallest to largest for Cone Size and Population (ASes that are lower in size and population are likely to have lesser collateral), while the frequency ordering is ASes with highest frequency to lowest frequency.}
  \label{fig:blocking-frequency}
\end{figure}

Figure~\ref{fig:blocking-frequency} compares how many proxies observed in real-world enumeration will be blocked if ASes are blocked in the order of different network-level strategies: frequency (most frequent $\rightarrow$ least frequent), AS population from APNIC~\cite{stats-apnic,salamatian2024s} (least populated $\rightarrow$ most populated) and AS cone sizes from CAIDA's AS-Rank dataset~\cite{as-rank} (smallest cone size $\rightarrow$ largest cone size). 

Blocking ASNs by observed proxy frequency is very effective because of Snowflake's concentrated nature: blocking the most frequent AS already covers about 30\% of all observed proxies, and the top 5\% of observed ASes by frequency cover more than 50\% of all proxies. By contrast, most observed proxies are deployed in ASes with large population sizes and cone sizes, as shown in Figure~\ref{fig:blocking-frequency}. This is expected since Snowflakes scales through volunteers who are connected to popular residential and access ASes. Therefore, attackers hoping to significantly block proxies will need to target networks with large user populations, which could lead to higher collateral for peer to peer applications such as gaming and file sharing. 

\section{Real-World Blocking Attack: Blocklist Stability}
\label{sec:app-real-world-blocklist-stability}
\begin{figure}[!t]
  \includegraphics[width=0.99\linewidth]{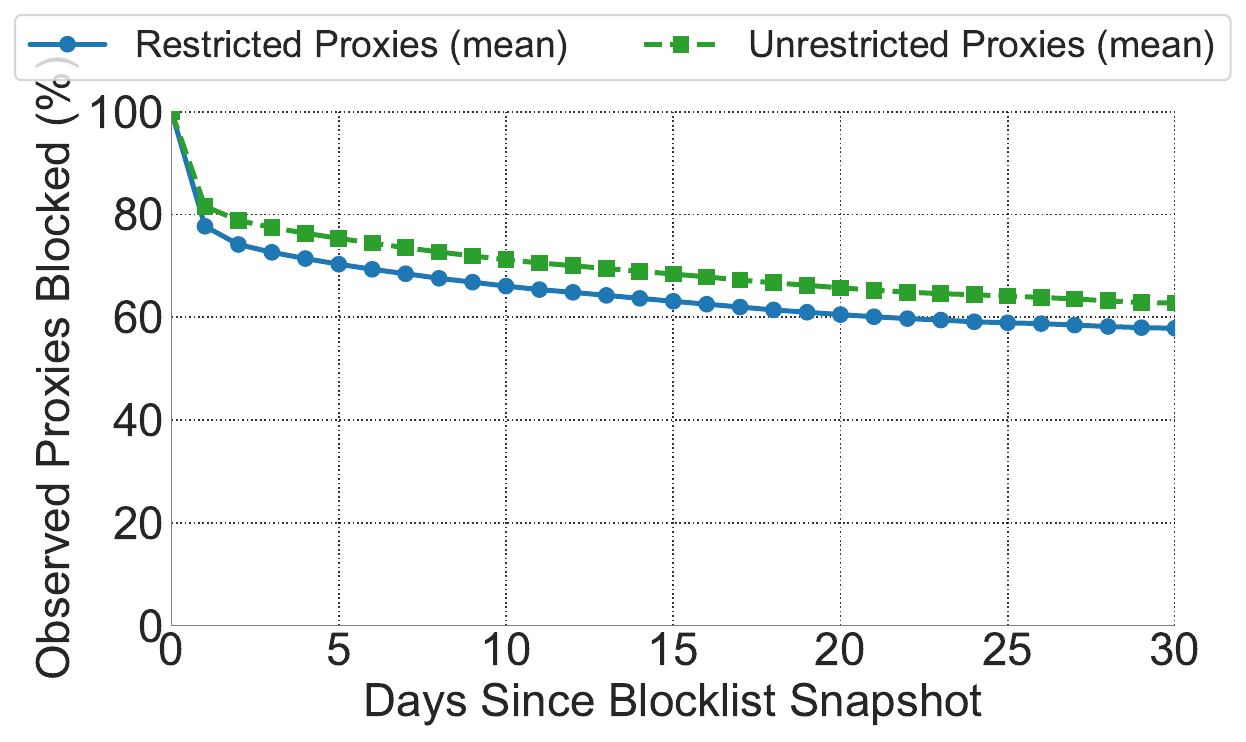}
  \caption{Decay of a frozen IP blocklist over time. An IP blocklist constructed from all real-world enumeration results observed up to a given snapshot day initially covers 77.7\% of the next day's active restricted proxies and 81.6\% of the next day's active unrestricted proxies, but this coverage declines steadily as the blocklist becomes stale.}
  \label{fig:blocking-stability}
\end{figure}

Figure~\ref{fig:blocking-stability} evaluates the shelf life of an IP blocklist after enumeration stops. For each snapshot day $d$, we freeze the blocklist containing all proxy IP hashes observed up to and including day $d$, and then measure what fraction of active proxies on later days $>d$ remain covered without any additional enumeration. The blocklist remains highly effective in the short-term: it covers $\approx$ 80\% of the next day's observed proxies. However, coverage falls slowly over time due to churn, reducing to $\approx$65\% after a week. Snowflake proxies, particularly lightweight ones in the restricted NAT set, are dynamic enough that a censor cannot enumerate once and block forever. Maintaining effective IP-blocking requires continuous probing. 

\section{Simulation Attack: 30m Default Connection Duration Results}
\label{sec:30m-default}

\begin{figure}[!t]
  \includegraphics[width=\linewidth]{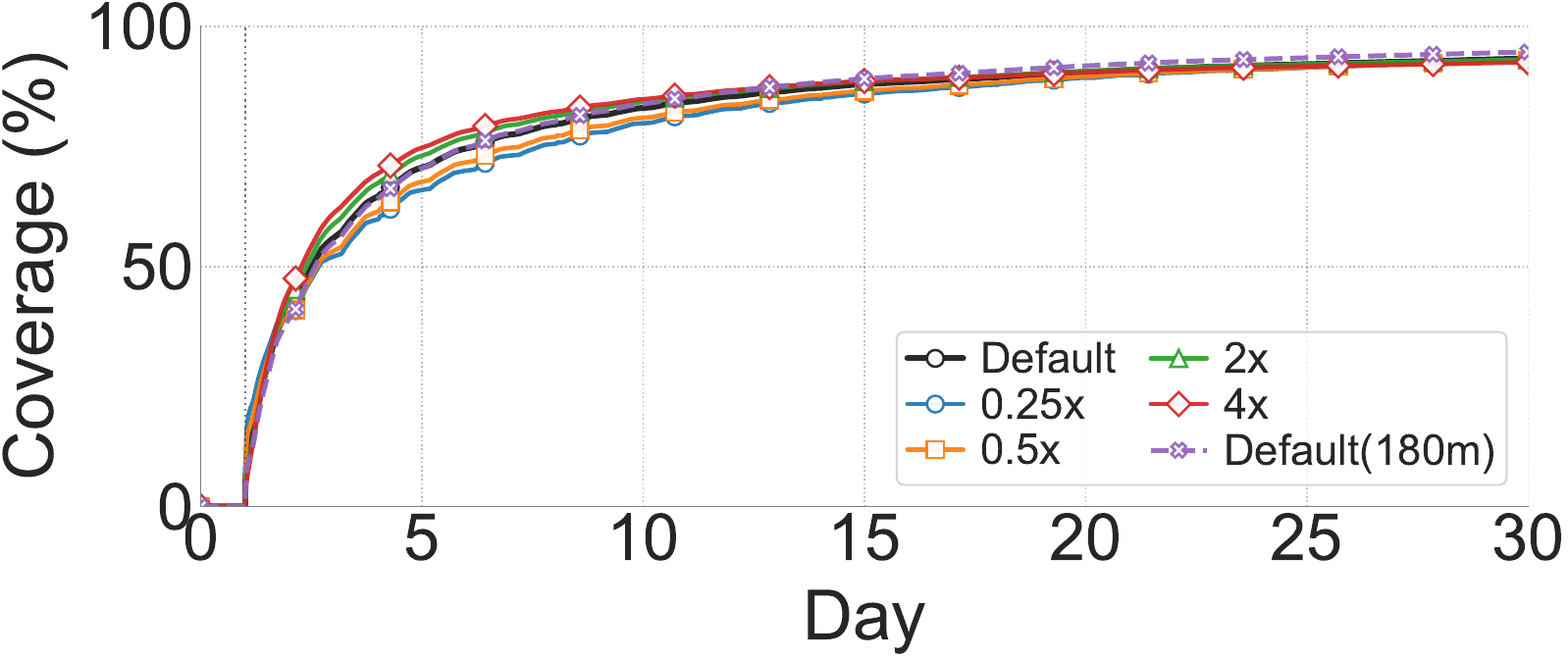}
  \caption{\textmd{Connection duration set to be 30 minutes, the comparison of different churn rate shows that low churn brings less coverage, which aligns with the previous experiments}}
  \label{fig:30m-churn-rate}
\end{figure}

\begin{figure}[!t]
  \includegraphics[width=\linewidth]{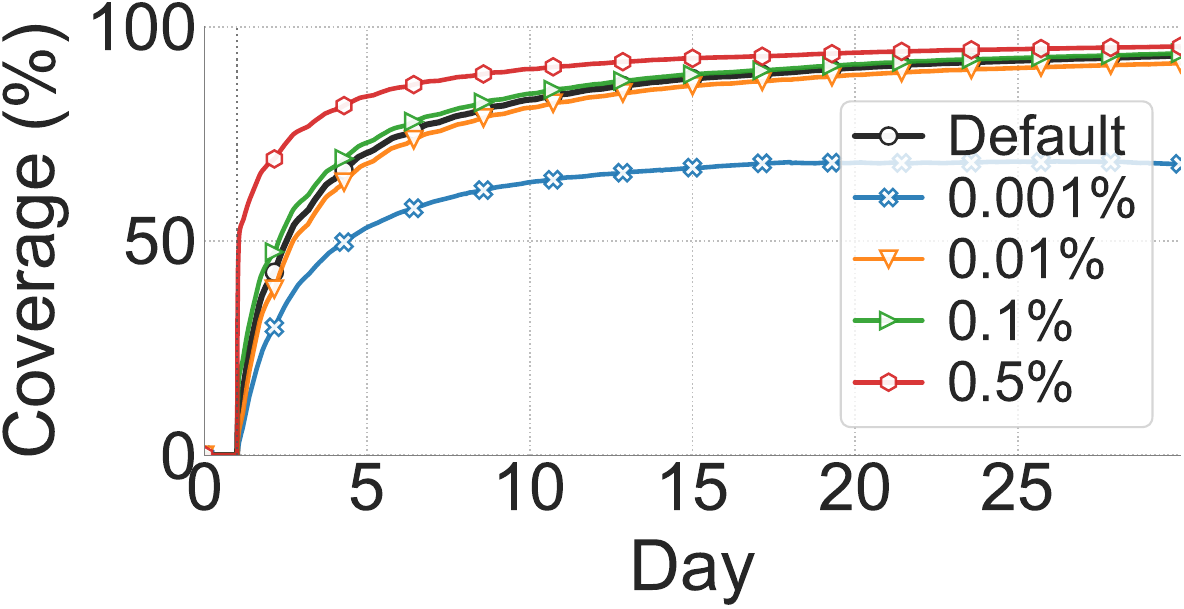}
  \caption{\textmd{Connection duration set to be 30 minutes, the comparison of different attacker fraction shows that fewer attackers cover less, which aligns with the previous experiments}}
  \label{fig:30m-attacker-count}
\end{figure}

The average duration of a Snowflake client--proxy connection is not available in public metrics. Our main experiments use a default connection duration of 180 minutes, calibrated against the published Snowflake client population~\cite{snowflake-clients} to maintain a plausible number of concurrent clients in the simulation. Here, we complement this analysis by reproducing the key attacker-fraction and churn-rate experiments using a mean connection duration of 30 minutes while holding all other parameters fixed.

Figures~\ref{fig:30m-churn-rate} and~\ref{fig:30m-attacker-count} present the enumeration results. Reducing the baseline connection duration to 30 minutes does not change the qualitative relationships between the experimental parameters and enumeration coverage. Increasing the attacker fraction increases coverage. Higher churn also produces slightly higher cumulative coverage because it introduces new proxies with no active clients; these proxies receive high priority in the broker heap and are therefore readily exposed to attackers. These trends are consistent with the results obtained using the 180-minute baseline.

\begin{figure*}[t]
    \centering

        \begin{subfigure}[t]{0.4\textwidth}
        \centering
        \includegraphics[width=\linewidth]{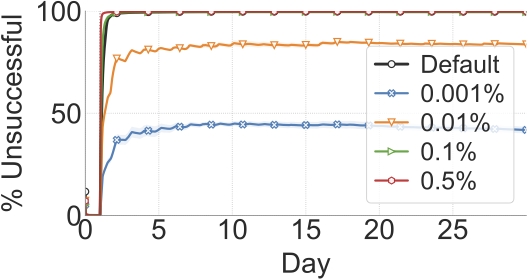}
        \vspace{0.25em}

        \includegraphics[width=\linewidth]{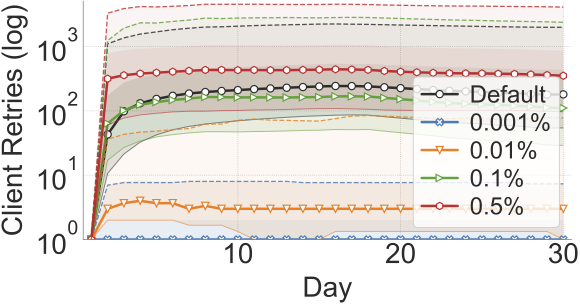}
        \caption{Attacker Fraction}
        \label{fig:default-30m-sim-block-attacker}
    \end{subfigure}
    \begin{subfigure}[t]{0.4\textwidth}
        \centering
        \includegraphics[width=\linewidth]{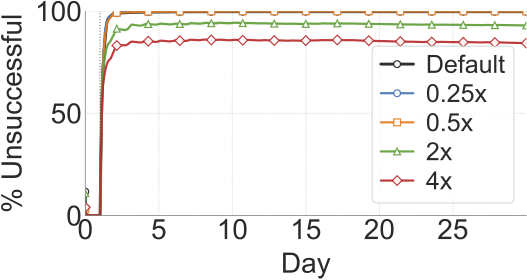}
        \vspace{0.25em}

        \includegraphics[width=\linewidth]{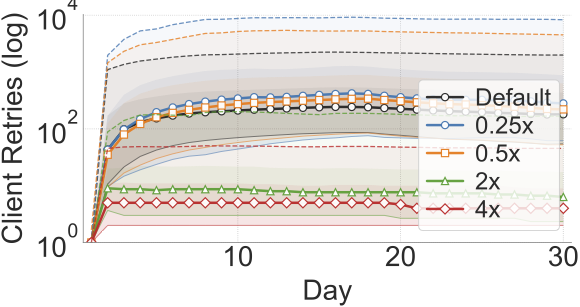}
        \caption{Churn Rate}
        \label{fig:default-30m-sim-block-churn}
    \end{subfigure}

    \caption{When the default connection duration is set to be 30 minutes, the effect of churn rate change and attacker count change is the same as our default connection duration setting to be 180 minutes.}
    \label{fig:default-30m-comparison}
\end{figure*}

Figure~\ref{fig:default-30m-comparison} shows that the blocking results also remain consistent with the baseline: increasing the attacker fraction or reducing proxy churn increases both the fraction of unsuccessful rendezvous attempts and the client retry burden. However, the absolute blocking impact is substantially greater with the 30-minute baseline. Under the default attacker and churn settings, the cumulative unsuccessful-request rate approaches 100\% before day 5, and the mean retry count among requests requiring at least one retry exceeds 100. Shorter connections return clients and proxies to the rendezvous process more frequently, increasing competition for unblocked proxies and causing clients to encounter blocked proxies repeatedly. Thus, connection duration substantially affects the magnitude of blocking harm, but does not change the direction of the effects associated with attacker scale and proxy churn.

\section{Simulation Enumeration Attack: Proxy-Type Results}
\label{sec:app-simulation-enumeration-results}
\begin{figure*}[t]
    \centering

    \includegraphics[width=\textwidth]{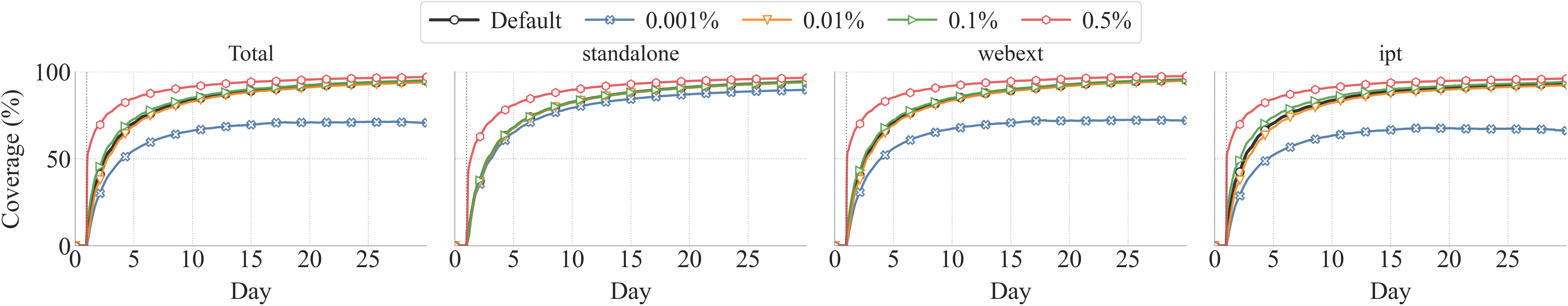}
    \caption*{(a) Attacker Fraction}

    \vspace{0.8em}

    \includegraphics[width=\textwidth]{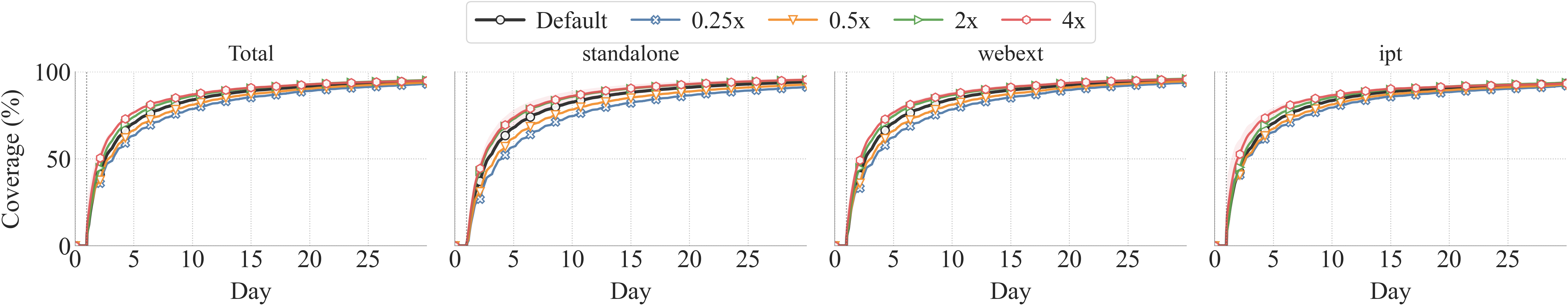}
    \caption*{(b) Churn Rate}

    \vspace{0.8em}

    \includegraphics[width=\textwidth]{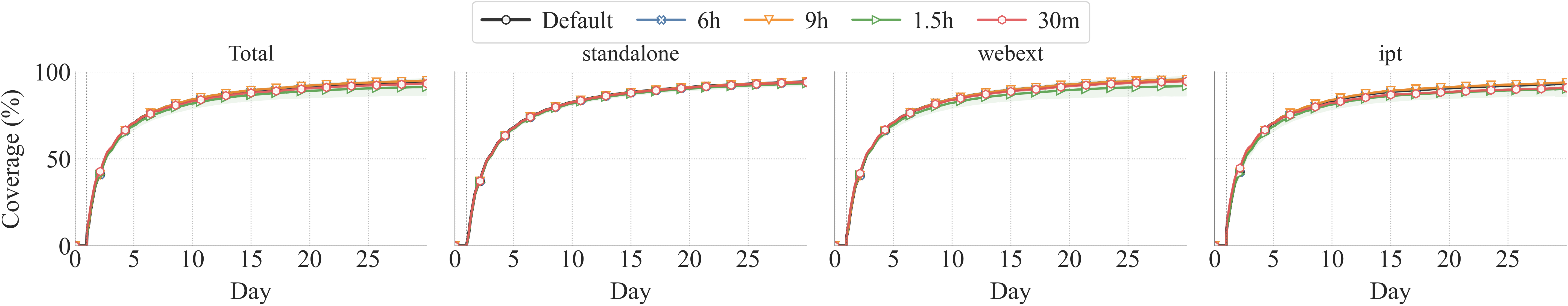}
    \caption*{(c) Connection Duration}

    \caption{Cumulative enumeration coverage by Snowflake proxy type under different simulation factor. Each panel varies one factor while holding all others at their default settings, and reports coverage separately for \texttt{Standalone}, \texttt{WebExt}, and \texttt{IPT} proxies. Our findings are similar across all proxy types.}
    \label{fig:appendix-simulation-enumeration-proxy-types}
\end{figure*}

Figure~\ref{fig:appendix-simulation-enumeration-proxy-types} shows that our enumeration findings generalize across Snowflake proxy types. Across most settings, \texttt{Standalone}, \texttt{WebExt}, and \texttt{IPT} proxies exhibit similar qualitative trends: attacker scale increases coverage, while the other system factors have smaller effects on cumulative enumeration. The main exception is the 0.001\% attacker setting for \texttt{Standalone} proxies. Because the \texttt{Standalone} pool is much smaller than the \texttt{WebExt} and \texttt{IPT} pools, even two attackers can enumerate a substantial fraction of \texttt{Standalone} proxies, whereas the same attacker scale achieves much lower coverage for the larger proxy classes.

\section{Simulation Blocking Attack: Client NAT-Type Results}
\label{app:app-simulation-blocking-results}
\begin{figure*}[t]
    \centering
    \begin{subfigure}[t]{0.49\textwidth}
        \centering
        \includegraphics[width=\linewidth]{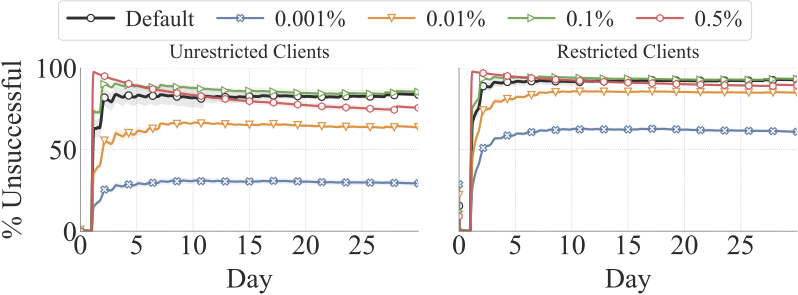}
        \vspace{0.25em}

        \includegraphics[width=\linewidth]{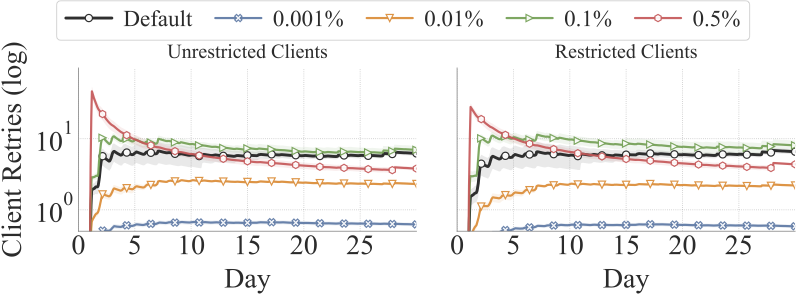}
        \caption{Attacker Fraction}
        \label{fig:app-sim-block-nat-attacker}
    \end{subfigure}\hfill
    \begin{subfigure}[t]{0.49\textwidth}
        \centering
        \includegraphics[width=\linewidth]{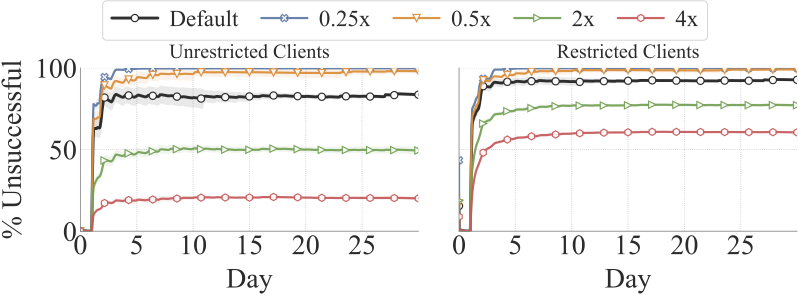}
        \vspace{0.25em}

        \includegraphics[width=\linewidth]{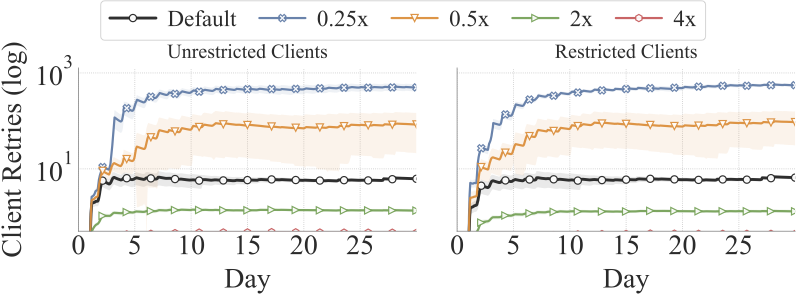}
        \caption{Churn Rate}
        \label{fig:app-sim-block-nat-churn}
    \end{subfigure}

    \vspace{0.75em}
    
    \begin{subfigure}[t]{0.49\textwidth}
        \centering
        \includegraphics[width=\linewidth]{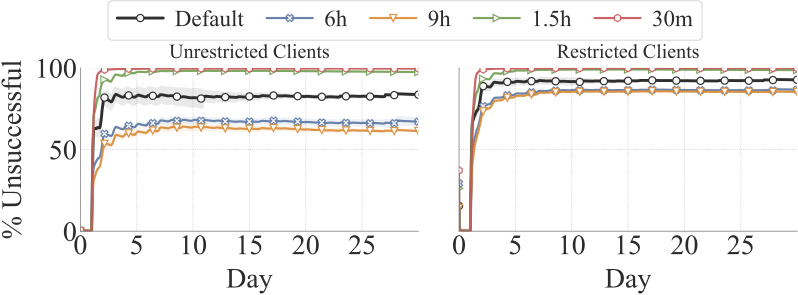}
        \vspace{0.25em}

        \includegraphics[width=\linewidth]{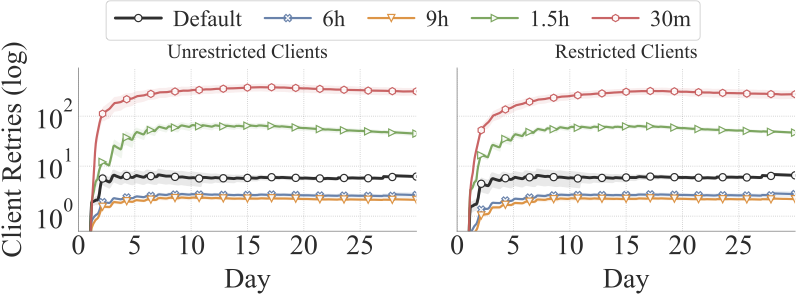}
        \caption{Connection Duration}
        \label{fig:app-sim-block-nat-connection}
    \end{subfigure}

    \vspace{0.75em}

    \caption{Blocking results by client NAT type across simulation factors. For each setting, : the top plot shows the cumulative fraction of rendezvous attempts up to hour $t$ that are unsuccessful and led to a retry, split by client NAT type. The bottom plot shows the corresponding mean retry burden among requests that required at least one retry.}
    \label{fig:appendix-simulation-blocking-client-nat}
\end{figure*}

Figure~\ref{fig:appendix-simulation-blocking-client-nat} shows that blocking attacks affect clients with restricted NAT types more than clients with unrestricted NAT types. Across all settings, restricted clients generally experience both more unsuccessful requests and a higher retry burden compared to unrestricted clients. In our investigations, we observe that unrestricted clients can match with the larger and more dynamic restricted-proxy pool, which is harder for attackers to enumerate and block. In contrast, restricted clients only use the unrestricted proxy pool, which is smaller and easily enumerated.

\end{document}